\documentclass[]{aastex631}

\usepackage{natbib}
\usepackage{xcolor}
\newcommand{\eps}[1]{\mbox{log~$\epsilon$(#1)}} 
\newcommand\species[2]{#1 {\sc #2}}
\newcommand\iso[2]{$^{\rm #1}$#2}

\def\eg{\mbox{e.g.}}
\def\etal{\mbox{\rm et al.}}
\def\teff{\mbox{$T_{\rm eff}$}}
\def\logg{\mbox{log({\it g})}}
\def\kmsec{\mbox{km~s$^{\rm -1}$}}

\def\vmicro{$V_{mic}$}
\def\vsini{$V$sin($i$)}

\def\he{\mbox{He {\sc i}}}
\def\rwhe{log($RW_{\rm He}$)}

\shorttitle{HPF Chemical Compositions}
\shortauthors{Sneden et al.}

\begin{document}

\title{THE ACTIVE CHROMOSPHERES OF LITHIUM-RICH RED GIANT STARS\footnote{
Based on observations obtained with the Hobby-Eberly Telescope, which is a 
joint project of the University of Texas at Austin, the Pennsylvania State 
University, Ludwig-Maximilians-Universit{\" a}t M{\" u}nchen, and 
Georg-August-Universit{\" a}t G{\" o}ttingen.}}

\correspondingauthor{Christopher Sneden}
\email{chris@verdi.as.utexas.edu}

\author[0000-0002-3456-5929]{Christopher Sneden}
\affiliation{Department of Astronomy and McDonald Observatory,
             The University of Texas, Austin, TX 78712, USA}
\author[0000-0002-2516-1949]{Melike Af\c{s}ar}
\affiliation{Department of Astronomy and Space Sciences,
             Ege University, 35100 Bornova, \.{I}zmir, Turkey}
\affiliation{Department of Astronomy and McDonald Observatory,
             The University of Texas, Austin, TX 78712, USA}
\author[0000-0002-4413-4401]{Zeynep Bozkurt}
\affiliation{Department of Astronomy and Space Sciences,
             Ege University, 35100 Bornova, \.{I}zmir, Turkey}
\author[0000-0002-6904-359X]{Monika Adam{\' o}w}
\affiliation{Center for AstroPhysical Surveys, National Center for 
             Supercomputing Applications, Urbana, IL 61801, USA}
\author[0000-0002-4282-605X]{Anohita Mallick}
\affiliation{Indian Institute of Astrophysics, Bangalore – 560034, India}
\author[0000-0001-9246-9743]{Bacham E. Reddy}
\affiliation{Indian Institute of Astrophysics, Bangalore – 560034, India}
\author[0000-0001-9165-8905]{Steven Janowiecki}
\affiliation{Hobby Eberly Telescope,
             The University of Texas, Austin, TX 78712, USA}
\author[0000-0001-9596-7983]{Suvrath Mahadevan}
\affiliation{Department of Astronomy \& Astrophysics, The Pennsylvania 
             State University, University Park, PA 16803, USA}
\affiliation{Center for Exoplanets and Habitable Worlds, The Pennsylvania 
             State University, University Park, PA 16803, USA}
\author[0000-0003-2649-2288]{Brendan P. Bowler}
\affiliation{Department of Astronomy and McDonald Observatory,
             The University of Texas, Austin, TX 78712, USA}
\author[0000-0002-1423-2174]{Keith Hawkins}
\affiliation{Department of Astronomy and McDonald Observatory,
             The University of Texas, Austin, TX 78712, USA}
\author{Karin Lind}
\affiliation{Department of Astronomy, Stockholm University, AlbaNova 
             University Centre, SE-106 91 Stockholm, Sweden}
\author[0000-0002-8985-8489]{Andrea K. Dupree}
\affiliation{Center for Astrophysics/Harvard \& Smithsonian,
             Cambridge, MA 02138-1516, USA}
\author[0000-0001-8720-5612]{Joe P. Ninan}
\affiliation{Department of Astronomy \& Astrophysics, The Pennsylvania 
             State University, University Park, PA 16803, USA}
\affiliation{Center for Exoplanets and Habitable Worlds, The Pennsylvania 
             State University, University Park, PA 16803, USA}
\author[0000-0002-7112-2086]{Neel Nagarajan}
\affiliation{Department of Astronomy and McDonald Observatory,
             The University of Texas, Austin, TX 78712, USA}
\author{Gamze B\"{o}cek Topcu}
\affiliation{Department of Astronomy and Space Sciences,
             Ege University, 35100 Bornova, \.{I}zmir, Turkey}
\author[0000-0001-8499-2892]{Cynthia S. Froning}
\affiliation{Department of Astronomy and McDonald Observatory,
             The University of Texas, Austin, TX 78712, USA}
\author[0000-0003-4384-7220]{Chad F. Bender}
\affiliation{Steward Observatory,
             University of Arizona, Tucson, AZ 85721, USA}
\author[0000-0002-4788-8858]{Ryan Terrien}
\affiliation{Department of Physics and Astronomy, Carleton College,
             Northfield, MN 55057, USA}
\author[0000-0002-4289-7958]{Lawrence W. Ramsey}
\affiliation{Department of Astronomy \& Astrophysics, The Pennsylvania 
             State University, University Park, PA 16803, USA}
\affiliation{Center for Exoplanets and Habitable Worlds, The Pennsylvania 
             State University, University Park, PA 16803, USA}
\author[0000-0001-7875-6391]{Gregory N. Mace}
\affiliation{Department of Astronomy and McDonald Observatory,
             The University of Texas, Austin, TX 78712, USA}

\begin{abstract}

We have gathered near-infrared $zyJ$-band high resolution spectra of nearly
300 field red giant stars with known lithium abundances in order to survey
their \species{He}{i} $\lambda$10830 absorption strengths. 
This transition is an indicator of chromospheric activity and/or mass loss
in red giants.
The majority of stars in our sample reside in the red clump or red 
horizontal branch based on their $V-J,M_V$ color-magnitude diagram and 
their Gaia \teff, \logg\ values.
Most of our target stars are Li-poor in the sense of having normally
low Li abundances, defined here as \eps{Li}~$<$~1.25.
Over 90\% of these Li-poor stars have weak $\lambda$10830 features.
But more than half of the 83 Li-rich stars (\eps{Li}~$>$~1.25) have strong
$\lambda$10830 absorptions.
These large $\lambda$10830 lines signal excess chromospheric activity
in Li-rich stars; there is almost no indication of significant mass loss.
The Li-rich giants also may have a higher binary fraction than do 
Li-poor stars, based on their astrometric data.
It appears likely that both residence on the horizontal branch and present or 
past binary interaction play roles in the significant Li-He connection 
established in this survey.

\end{abstract}

\section{INTRODUCTION}\label{intro}

Lithium is easily destroyed in stellar interiors as part of the proton-proton
fusion cycles.
At modest fusion temperatures, T~$\geq$~2.5$\times$10$^6$~K, the reaction
\iso{7}{Li}($p,\alpha$)$\rightarrow$\iso{4}{He} efficiently cleans out
lithium in interior regions.
Then basic stellar evolution computations (\eg, \citealt{iben67a}) predict
that for metal-rich stars the deepening convective envelopes during subgiant
and first-ascent red giant evolution will dilute the initial surface Li
abundances by factors approaching $\sim$60.
In this way the stellar age-zero Li contents
(\eps{Li}~$\simeq$~3.3)\footnote{
We adopt the standard spectroscopic notation
\citep{wallerstein59} that for elements A and B,
[A/B] $\equiv$ log$_{\rm 10}$(N$_{\rm A}$/N$_{\rm B}$)$_{\star}$ $-$
log$_{\rm 10}$(N$_{\rm A}$/N$_{\rm B}$)$_{\odot}$.
We use the definition
\eps{A} $\equiv$ log$_{\rm 10}$(N$_{\rm A}$/N$_{\rm H}$) + 12.0, and
equate metallicity with the stellar [Fe/H] value.}
that are observed in warm main sequence stars (\citealt{randich20},
\citealt{romano21}) and in the primordial solar system (\citealt{lodders21})
will be substantially diminished, leaving normal red giants with
\eps{Li}~$\sim$~1.5.
But few evolved stars have that much surface Li.
Decades ago surveys of Li in red giants (\eg, \citealt{Wallerstein69}, 
\citealt{lambert80}, \citealt{brown89b})
established that nearly all red giants have \eps{Li}~$<$~1.0, and abundance 
upper limits are common.
This is consistent with the small Li abundances in cool 
main-sequence stars (dating back to \citealt{Wallerstein69} and
 \citealt{boesgaard86,boesgaard87}), those with \teff~$<$~6500~K.
Slow circulation currents during long main sequence lifetimes deplete the
surface Li abundances before stars evolve to become giants.
The very low Li content of normal red giant stars is no mystery.

However, about 1\% of red giants have unexpectedly strong \species{Li}{i} 
6707~\AA\ absorption lines, often leading to Li abundances that are much 
greater than the maximum values in main-sequence stars.
The first serendipitous discovery of a Li-rich giant \citep{wallerstein82}
has been followed by many other detections; now more than
100 relatively bright Li-rich evolved stars have been cataloged.
The origin of this phenomenon is not clear.
The best ``interior'' hypothesis is the so-called beryllium transport
mechanism \citep{cameron71}, in which the reactions
$^3$He($\alpha,\gamma$)$^7$Be(e$^-$,$\nu$)$^7$Li create Li, which then
must be  quickly convected to the surface before it can decay by
capturing a proton as described above.
The most prominent ``exterior'' hypothesis invokes terrestrial or hot Jupiter
planet engulfment by stars \citep{alexander67} as they bloat themselves at
post-main-sequence life stages.

The observational clues are mixed, but recent evidence appears to favor
the interior production and mixing explanation.
The \cite{adamow14} spectroscopic red giant survey found that their
four most Li-rich giants (log $\epsilon$(Li)~$>$~1.5) either have low-mass
companions or significant radial velocity variations, suggesting that
companions play roles in the Li-enrichment processes.
But \cite{singh19b} used asteroseismic and spectroscopic
analyses of more than 10,000 giants to show that nearly
all of the very Li-rich giants in their survey are in the He-core
burning red clump (RC) stage.
\cite{casey19} combined these ideas to suggest that tidal interactions
between (primarily) He-core-burning RC giants and binary companions
can induce internal mixing to bring newly-created
Li to the stellar surfaces for brief ($\sim$10$^6$ yr) episodes.
\cite{singh21} have now shown that very high Li abundances are
associated with ``young'' He-core-burning giants that have just recently
arrived on the clump following the RGB-tip He-flash.
Finally, \cite{deepak20} argue that the lack of correlation of red giant Li 
abundances with rotational velocity means that mergers and tidal interactions
are unassociated with Li enhancements.
However, the large survey by \cite{mallick22} for IR excess among RC 
stars shows that Li-rich giants with IR excess are more likely to be 
fast rotators compared to Li-normal giants.

\begin{figure}                                                
\epsscale{0.70}                                               
\plotone{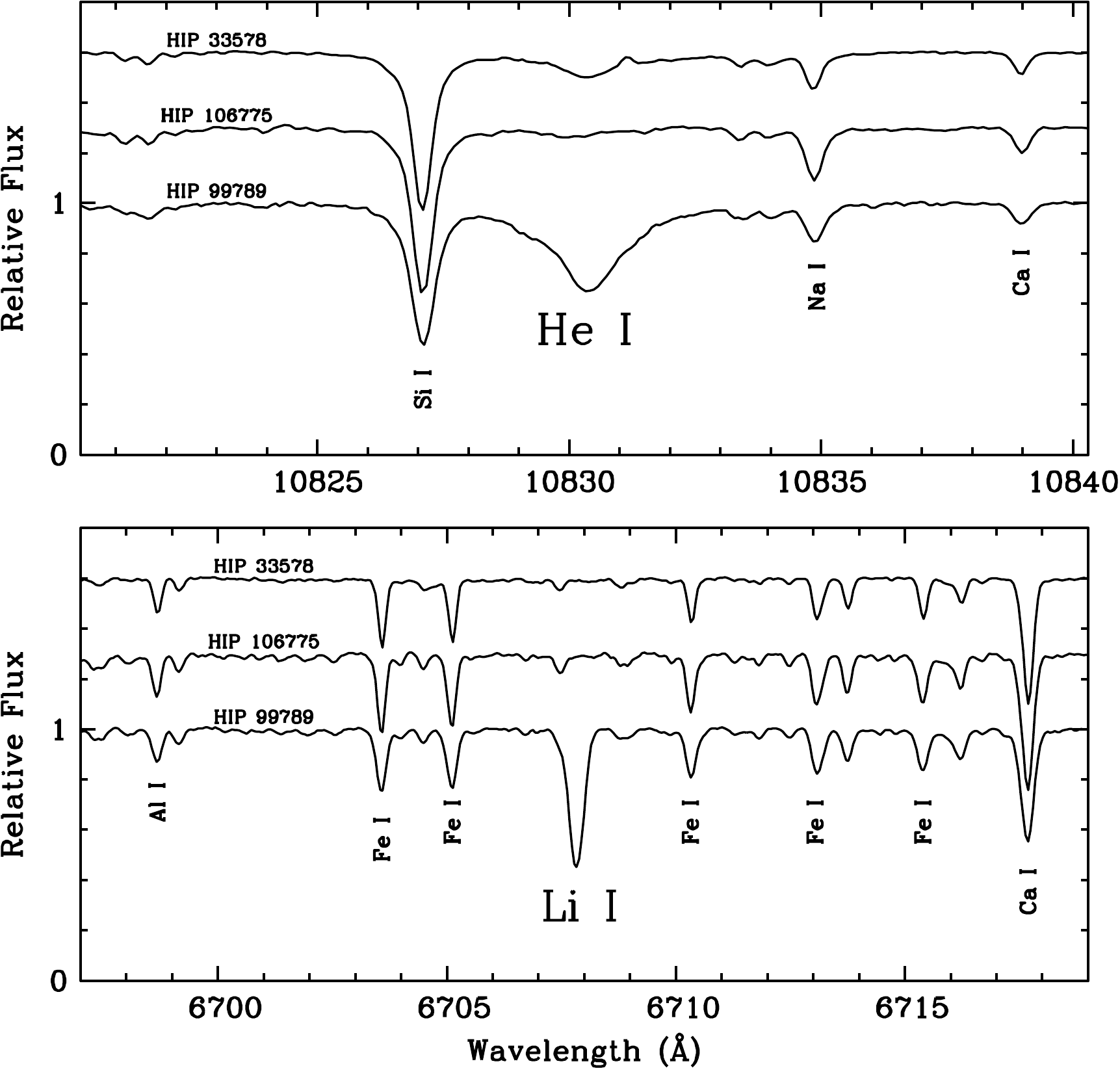}                                        
\caption{\label{fig1}                                     
\footnotesize                                                 
   Top panel: He I 10830~\AA\ spectra of three red horizontal branch stars 
   \citep{sneden21}.
   Bottom panel: the Li I 6707~\AA\ resonance feature in these 
   stars (\citealt{afsar18a}, Bozkurt et al. in preparation).
}                                                             
\end{figure}

Recently \cite{sneden21} used The Habitable Zone Planet Finder (hereafter HPF)
on the Hobby-Eberly Telescope (HET) to conduct a $zyJ$ high resolution
spectroscopic investigation of some red horizontal branch (RHB), or
secondary RC field giants.
That study discovered a strong \species{He}{i} 10830~\AA\ line in the Li-rich 
RHB star HIP~99789.
We illustrate the Li and He spectral features in HIP~99789 and two other
RHB's in Figure~\ref{fig1}.
The lower excitation energy of the \species{He}{i} 10830~\AA\ transition 
is 19.8~eV; thus this line cannot be formed in the cool photospheric layers 
that yield \species{Li}{i} resonance line absorption. 
Strong $\lambda$10830 transitions in cool giant stars signal either excess
chromospheric activity or outer atmospheric wind outflow.   
The notion that \species{He}{i} absorption might be a new clue to the        
Li-rich giant phenomenon has led us to a survey of \species{He}{i} 
transitions in Li-rich and Li-poor red giants that we report in this paper. 

In this survey the first goal was to discover the fraction of Li-rich     
giants that have large $\lambda$10830 features.
Then we looked for correlations between \species{He}{i} absorption and 
other stellar parameters to help clarify the roles of Galactic population,
rotation, chromospheric activity, and stellar/planetary companions on the 
appearance of strong \species{He}{i} absorption and/or excess Li in giant 
stars.  
In \S\ref{targets} we describe the stellar sample, and in 
\S\ref{obsred} we outline the observations and spectroscopic reduction steps.
\S\ref{analysis} presents the process we used to isolate the $\lambda$10830
chromospheric lines from the photospheric spectra, to measure their 
equivalent widths, and to estimate stellar line broadening.
We give the basic results in \S\ref{results}, using Li-He correlations to 
define empirical regions of Li abundance and He absorption line strength, 
and suggest that in all but one case the He lines indicate chromospheric 
activity rather than mass loss.
Correlations of He strengths with space and possible binary motions are
presented in \S\ref{move}, and we summarize our work in \S\ref{conclusions}.

\vspace*{0.2in} \section{THE STELLAR SAMPLE}\label{targets} 

Lithium abundances in evolved stars range over at least five orders of 
magnitude, but almost all red giants have \eps{Li}~$<$~1.5.
Therefore we initially adopted \eps{Li}~=~1.5 as the upper limit for evolved 
stars that have suffered the Li destruction, envelope mixing, and 
surface dilution expected during ordinary stellar evolution.
For simplicity in this paper we label all such normal Li-depleted red giants
as ``Li-poor'' stars.
Evolved red giants with Li abundances above this limit are given the blanket
label ``Li-rich''.
This empirical definition will be reconsidered in \S\ref{results}.

Li-rich giant stars have been reported in individual spectroscopic 
studies (\eg, \citealt{luck82}, \citealt{wallerstein82}, 
\citealt{balachandran00}), 
and in large-sample Li discovery surveys (\eg, \citealt{brown89b}, 
\citealt{kumar11}, \citealt{adamow14}, \citealt{casey19}, \citealt{deepak19}).
We searched the growing literature on this phenomenon and developed a
target list based on the following criteria.
\begin{enumerate}
\item Confirmed evolved-star status.  
In general this included stars that are cool (\teff~$\lesssim$~5500~K, 
or $B-V$~$\gtrsim$~0.8) and well off the main sequence (\logg~$<$~3.5,
or $M_V$~$<$~$+$1.0).
\item Sky location in the HET-accessible domain, which is set by its
declination limits: 
$-$10$^{\circ}$ $\lesssim$ $\delta$ $\lesssim$ $+$71$^{\circ}$.
\item Target brightness reasonable for HPF.
The apparent magnitude bright limit is $J$~$>$~3.0~mag, driven by the need to 
avoid HPF detector saturation.
The practical faint limit is $J$~$\sim$~13.0, to achieve signal-to-noise
$S/N$~$>$~30 in several hours total observing time.\footnote{
For a more complete discussion of HPF integration times see
https://psuastro.github.io/HPF/Exposure-Times/}
\end{enumerate}

For Li-poor stars we primarily used two large high resolution optical
spectroscopic surveys of field red giants.
\cite{adamow14} determined Li abundances for 348 stars as part of the 
Penn State-Toru\'{n} Centre for Astronomy Planet Search program;
see also atmospheric parameter and overall chemical abundance studies
by \cite{zielinski12} and \cite{niedzielski16}.
These G-K giants lie in the atmospheric parameter range
4000~K~$\lesssim$~\teff~$\lesssim$~5100~K, 
$-$3.0~$\lesssim$~$M_V$~$\lesssim$~3.0.

We also observed some red horizontal branch stars, alternatively called 
the secondary
red clump \citep{girardi98}, from the \cite{afsar18a} 340-star survey.
Li abundances for these stars will be published by Bozkurt \etal\ (in
preparation).
These two large samples have Li abundances derived in a uniform manner,
which is important for those stars with very weak \species{Li}{i} 
$\lambda$6707 features.

When early observations revealed that Li-rich stars can also be
rapidly-rotating red giants (\vsini~$\gtrsim$~7~\kmsec), we also added
some extra targets with substantial rotation from several literature
sources: \cite{bizyaev10}, \cite{carlberg12}, \cite{costa15}.
These papers provided some bright stars for relatively easy HPF spectrum
acquisition, but do not constitute a comprehensive list of rapidly rotating
G-K giants.

In Table~\ref{tab-stars} we list basic data for the program stars, and
the most important measured quantities in this study.
The original Li abundance papers used a variety of names for these stars.
But almost all of our survey red giants have identifications in the
Tycho-2 catalog \citep{hog00}\footnote{
https://www.cosmos.esa.int/web/hipparcos/tycho-2}.
We adopt these ``TYC'' designations throughout this paper when possible.
However, many discovery papers on Li-rich red giants have focused on
relatively bright stars, and have identified them with more traditional
names from the HD or BD catalogs.
Therefore we also include in Table~\ref{tab-stars} the HD numbers when
available, else BD numbers when possible.
There are 278 program stars entered in Table~\ref{tab-stars}.
Of this total, 76 stars are Li-rich according to the tentative definition
discussed above, 187 stars are Li-poor, and 15 stars have no Li abundance.

Some of the program stars have more than one published Li abundance.
Our work with the \species{He}{i} $\lambda$10830 transition does not require
detailed assessment of literature Li values, so for this paper we have cited a
single Li reference in Table~\ref{tab-stars} for each star.
We have not undertaken any renormalization of the abundances, and quote
\eps{Li} to just one significant digit in the table.

Table~\ref{tab-stars} also contains two additional sets
of stars that are special-purpose targets.
The first group are red giant members of NGC~7789.
This is an open cluster with intermediate-age ($t$~$\sim$~1.7~Gyr;
WEBDA, \citealt{mermilliod95})\footnote{
https://webda.physics.muni.cz/navigation.html}
and solar metallicity, \eg, \cite{overbeek15}, \cite{donor18}, \cite{carrera19},
\cite{casamiquela19}.
\cite{pilachowski86} discovered that NGC~7789 hosts one and possibly two
Li-rich giants, which made this cluster an attractive target for our
$\lambda$10830 survey.
We obtained spectra for 10 of Pilachowski's red giants, and we tabulate here
our $\lambda$10830 measurements for them, identifying them by cluster and
star name.
However, we defer discussion of these data to a more general study that
contains new optical high-resolution spectra for this cluster
(Nagarajan \etal, in preparation).

The second special target group are Kepler satellite field
giants drawn from those studied by \cite{takeda17,singh19b,singh21}.
These are identified in Table~\ref{tab-stars} by their Kepler Input Catalog
(KIC) numbers.
The Kepler field giants have detailed evolutionary state information obtained
through asteroseismology.
We have observed 13 of these stars, but our selection was heavily biased
toward those with known very high Li abundances.
We are now gathering $\lambda$10830 observations for a much larger Kepler
giant sample.
Therefore we have chosen to report here the He measurements already done,
but will defer discussion of them until the larger set of spectra is gathered
(Mallick \etal, in preparation).

\vspace*{0.2in}
\section{OBSERVATIONS AND REDUCTIONS}\label{obsred}
We gathered HET/HPF high$-$resolution $zyJ$ (8100$-$12750~\AA) spectra of 
the program stars for about a year beginning in April 2021.
HPF was designed to aid in the search for low-mass companions to M-dwarf 
stars, and it has unique design features that deliver extremely 
high-precision radial velocities (down to $\sim$1-2~m~s$^{-1}$).
This instrument has been described in detail by 
\cite{mahadevan12,mahadevan14}.\footnote{
see \texttt{https://hpf.psu.edu/} for additional instrument information.}
HPF was configured for our program to deliver spectral resolving power
$R$~$\equiv$~$\lambda/\Delta\lambda$~$\sim$~55,000.

In the $zyJ$ spectral region, telluric molecular blockage ranges from near 
zero to almost a complete blanket, varying significantly with wavelength.
The dominant telluric contaminators are H$_2$O bands, and their spectrum
blockage can change substantially on timescales of hours to nights.
Regular observations are obtained of rapidly-rotating hot stars that can
be used to cancel the telluric features in the program stars.
The HET's fixed altitude is an asset here: all stars have identical
air masses, thus it is not necessary to gather hot-star spectra to accompany
each target spectrum.
Divisor-star spectra adequate for our needs are normally gathered 1-2 times 
during nights with stable atmospheric conditions.

\begin{figure}
\epsscale{0.70}
\plotone{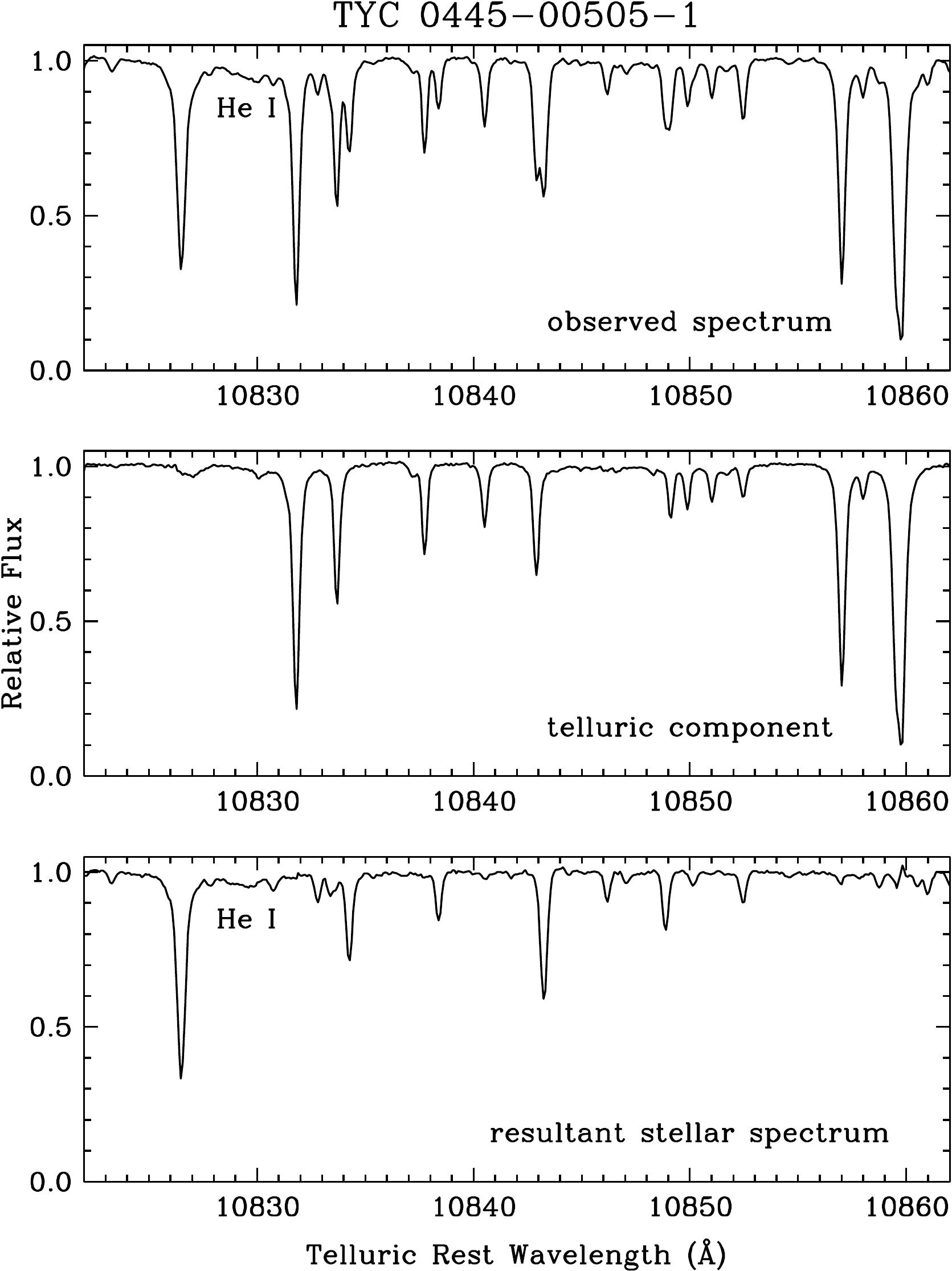}
\caption{\label{fig2}
\footnotesize
   An example of telluric absorption line cancellation near the 
   $\lambda$10830 transition.
   For this figure a program star with very high S/N but weak \he\
   absorption depth was chosen.
   The top panel contains the stellar+telluric spectrum, and the middle panel 
   has the spectrum of a warm (essentially featureless) star.
   The division of these two spectra produces the purely stellar program
   star spectrum in the bottom panel.
}
\end{figure}

HPF observations of target and telluric divisor stars were initially reduced
with the HPF pipeline code $Goldilocks$.\footnote{
https://github.com/grzeimann/Goldilocks\_Documentation}
This facility, run automatically within a day of data acquisition, processes 
raw HPF data frames into useful spectra by removing 
bias noise, correcting for instrumental nonlinearity, masking significant 
anomalous radiation events, calculating the slope/flux and 
variance image using the algorithms from the $pyhxrg$ module in the tool 
$HxRGproc$ \citep{ninan18}, and performing optimal extraction of the
28 spectral orders.  
HPF's wavelength scale in spectra reduced with $Goldilocks$ 
is extremely stable and accurate to levels far beyond that needed for
our analyses of the program star spectra.
For each stellar source, the $Goldilocks$ spectral files have several 
extensions, including ones for the target, the local night sky emission,
noise, and wavelength calibrations.

The rest of the reductions were performed with the IRAF\footnote{
http://iraf.noao.edu/} \citep{tody86,tody93} facility.
We used the \textit{disptrans} task to shift the $Goldilocks$ vacuum 
wavelengths to air wavelengths.
Sky subtraction and mating of target fluxes and wavelengths produced standard
echelle FITS spectra.
We then performed continuum normalization of the target spectra and
hot star spectra via the \textit{continuum} task and used the 
\textit{telluric} task interactively to divide out this contamination
from the target spectrum.
Figure~\ref{fig2} illustrates the large number of telluric features in
the HPF spectral order containing $\lambda$10830, and the necessity of
iterative division operations to produce optimal elimination of the tellurics.
A final continuum normalization was followed by merging into a single 
continuous spectrum with the \textit{scombine} task.

To correct observed stellar velocities to a rest wavelength scale 
accurate enough for our purposes, we first determined the velocity shifts
using IRAF $splot$ to measure the observed wavelengths of several strong
atomic lines in the spectral order containing the \species{He}{i}
$\lambda$10830 feature. 
Their average displacement from rest in velocity units provided a
shift to stellar rest velocity accurate to $\sim\pm$0.1~\kmsec.

\begin{figure}
\epsscale{0.80}
\includegraphics[angle=-90,scale=0.75]{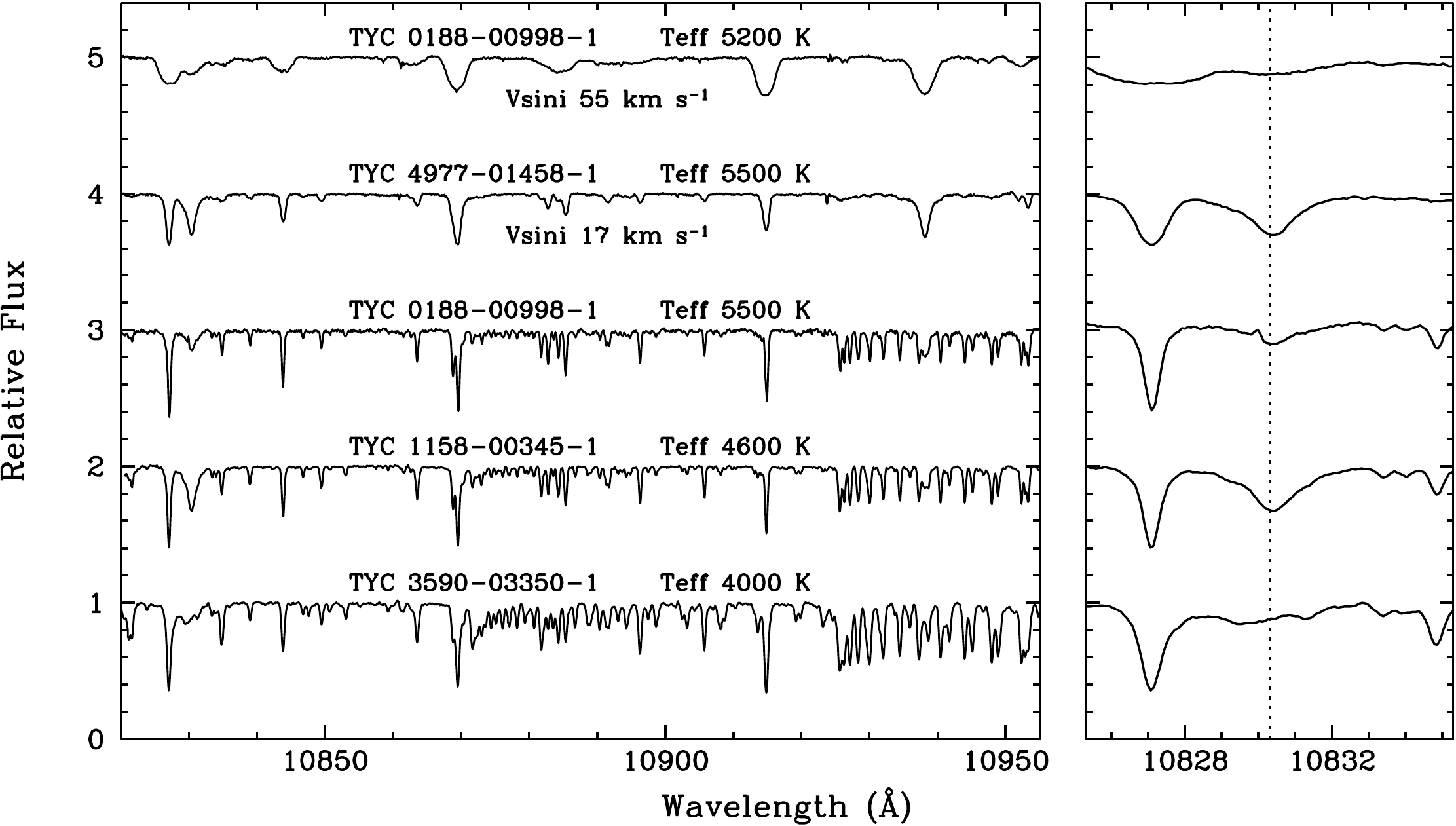}
\caption{\label{fig3}
\footnotesize
   Reduced and telluric-corrected spectra of seven program stars in the
   HPF spectral order that contains \species{He}{i} 10830.3~\AA.
   The left-hand panel shows the entire HPF order that contains the 
   $\lambda$10830 line.
   The right-hand panel zooms in to a 10~\AA\ interval centered on the
   $\lambda$10830 feature.
   The wavelength of the strongest two components of the \species{He}{i} 
   triplet is shown with a dotted line.
}
\end{figure}

In Figure~\ref{fig3} we display spectra of some typical program stars in 
the HPF order that contains the \species{He}{i} $\lambda$10830 feature.
The left-hand panel shows essentially the entire order, covering the 
spectral range $\sim$10820$-$10960~\AA.
Star TYC~3590-03350-1 is at the low-temperature end, and TYC~0188-00998-1
is at the high-temperature end of our sample.
Star TYC~1158-00345-1 has one of the larger Li abundances, \eps{Li}~=~3.1
(\citealt{kumar11} and references therein).
TYC~4977-01458-1 and TYC-0188-00998-1 have high rotational velocities,
and they have strong \species{He}{i} absorption lines, common among our 
program stars.
The right-hand panel shows just a small wavelength interval centered on the
$\lambda$10830 line.

\vspace*{0.2in}
\section{HE I $\lambda$10830 ANALYSIS}\label{analysis}

Red giants present rich absorption-line spectra in the $zyJ$ spectral domain.
But the present purpose is to investigate just the \he\ $\lambda$10830
transition.
This line in absorption connects the energy states $1s2s^3S$ (19.82~eV) 
and $1s2p^3P^o$ (20.96~eV), as illustrated in Figure~3 of \cite{preston22}.
Neither the lower nor the upper state has a permitted transition to the 
$1s^{2}$~$^{2}S$ ground state.
It is a hyperfine-split triplet, with components at 10829.09, 10830.25, and
10830.34~\AA, with transition probabilities $-$0.745, $-$0.268, and
$-$0.046 respectively in the NIST database 
\citep{kramida19,kramida19a}.\footnote{               
National Institute of Standards and Technology Atomic Spectra Database:     
https://www.nist.gov/pml/atomic-spectra-database}  

Appearance of the very high-excitation \species{He}{i} $\lambda$10830 line in 
cool red giants essentially signals chromospheric activity that is detached 
from the formation of the photospheric spectrum.  
Its line profile can be very broad, in some cases covering several \AA ngstroms.
In this small region there are several photospheric absorption lines,
including the very strong \species{Si}{i} 10827.1~\AA\ transition and
other weak-but-detectable ones:  \species{Ti}{i} 10827.9~\AA,
\species{Ca}{i} 10829.3~\AA, and a couple of significant CN lines.
In particular the \species{Ca}{i} line can significantly contaminate
the \species{He}{i} transition.
We wanted to make accurate estimates of the equivalent width ($EW$) of
only the $\lambda$10830 \species{He}{i} feature, and so we created
synthetic spectra to match and then cancel the surrounding photospheric
atomic and molecular transitions.

\subsection{Line List for Synthetic Spectra}\label{linelist}              

To generate line lists for the synthetic spectra, we used the $linemake$
facility \citep{placco21}\footnote{
https://github.com/vmplacco/linemake},
which begins with the \cite{kurucz11,kurucz18}\footnote{
http://kurucz.harvard.edu/linelists.html}
line compendium, and substitutes or adds line transition data generated in
recent laboratory studies mainly by the Wisconsin-Madison atomic physics group
(\citealt{denhartog21} and references therein) and by the Old Dominion
University molecular physics group (\eg, \citealt{brooke16} and references 
therein).  
Unfortunately there have been few laboratory atomic studies of transitions 
in the past couple of decades for the lines of interest near 10830~\AA.
Therefore we adopted the transition probabilities available in the NIST
database for lines of the most prominent species in this spectral region: 
\species{Na}{i}, \species{Mg}{i}, \species{Al}{i}, \species{Ca}{i}, 
\species{Fe}{i}, and \species{Sr}{ii}.

We tested this line list by generating synthetic spectra of the red giant
standard star Arcturus.
The observed spectrum was from \cite{hinkle05}. 
We adopted the model atmosphere parameters derived by \cite{ramirez11}:
\teff~= 4286~K, \logg~=~1.66, [Fe/H]~=~$-$0.52, \vmicro~=~1.74~\kmsec, as
well as the elemental abundance ratios given in that paper.
We produced an interpolated model with these values from the 
\cite{kurucz11,kurucz18} grid\footnote{
http://kurucz.harvard.edu/grids.html} 
using software developed by Andy McWilliam and Inese Ivans (private 
communication).
The synthetic spectra for Arcturus and all other stars in this paper
were generated using the current version of the LTE synthetic spectrum code 
MOOG \cite{sneden73}\footnote{
Available at http://www.as.utexas.edu/$~$chris/moog.html}.

Iterative comparison of observed and synthetic Arcturus spectra led to
adjustments in the transition probabilities and wavelengths of some transitions
in the line list.
In making these changes we did not alter any atomic or molecular line data
that were from NIST or recent laboratory analyses.
Finally, to verify the reasonableness of the final line list we also
checked its applicability to the spectra of the Sun and a couple of
program stars.
This procedure yielded an atomic/molecular line list for our
task of adequately matching synthetic and observed photospheric 
spectra of our red giant stars, but should not be confused with a list
that would be appropriate for detailed abundance analyses.

\subsection{Estimated Model Photospheres}\label{modatm}              

Nearly all of our program stars are Galactic disk red giants, having
metallicities almost entirely in the domain [Fe/H]~$\gtrsim$~$-$0.5.  
With the goal limited to adequate removal of the photospheric spectrum
near the \he\ $\lambda$10830 feature, we adopted effective temperatures
\teff\ from photometry and surface gravities \logg\ representative of
first-ascent, RC, and RHB giants at the chosen \teff\ values.

\begin{figure}                                                
\epsscale{0.70}                                               
\plotone{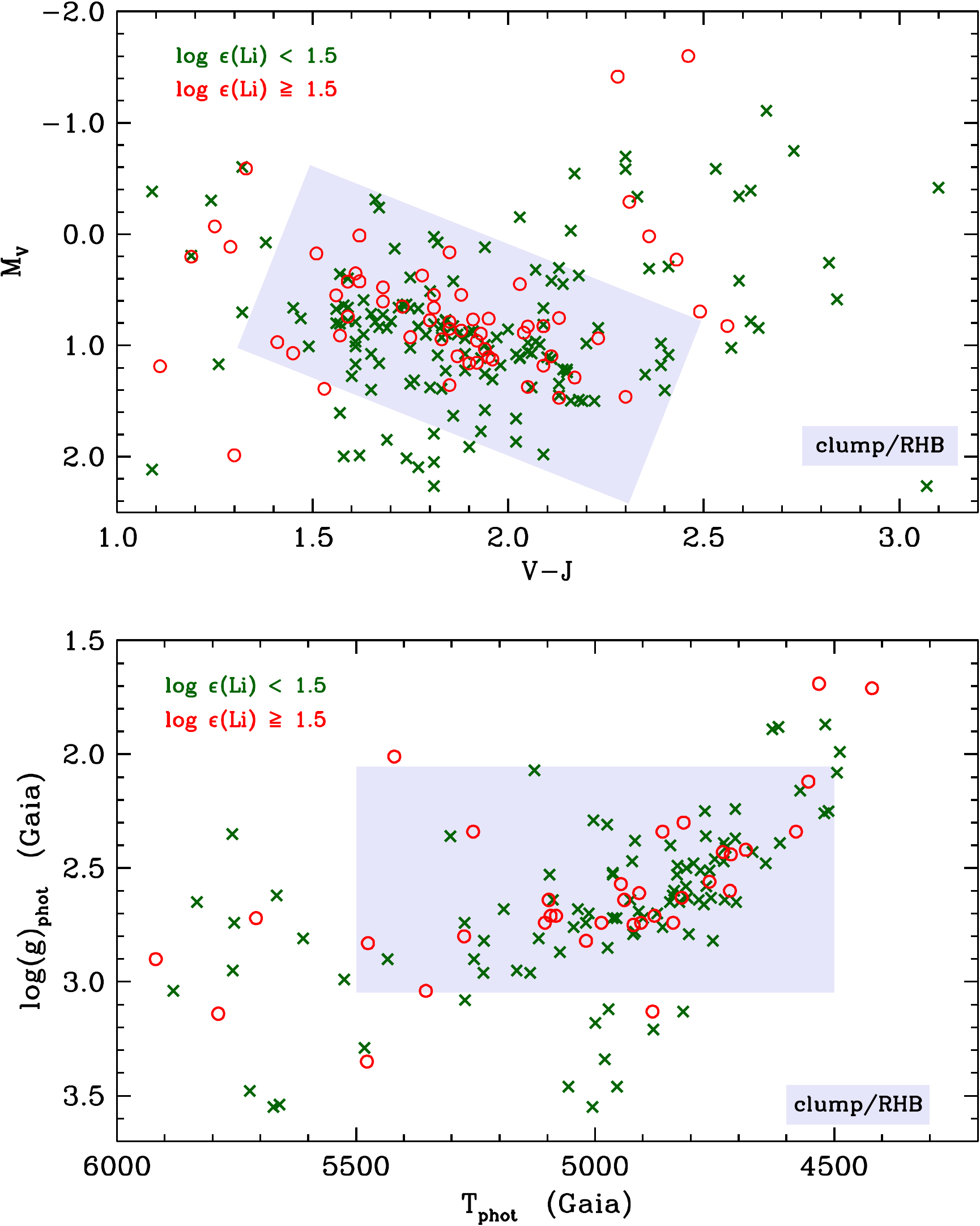}                                       
\caption{\label{fig4}                                    
\footnotesize                                                 
   Top panel: a color-magnitude diagram (CMD) for the targets.
   The plot boundaries have been set to show only the CMD area containing   
   the first-ascent red giants, the RC, and the RHB.
   The main sequence, most of the subgiant branch, and the luminous giants  
   and supergiants are outside of this CM area.             
   Different colors and symbols denote Li-rich and Li-poor stars, and a      
   freehand shaded region represents the He-fusing RC/RHB domain.        
   Bottom panel:  an HR diagram based on $Gaia$ photometric quantities.     
   Colors and symbols are the same as in the top panel.
}                                                             
\end{figure}

We first constructed a set of model stellar photospheres.
The desired parameter domain can be seen both in a color-magnitude diagram 
(CMD) and a \teff$-$\logg\ HR diagram of our targets.
From the magnitudes and parallaxes in Table~\ref{tab-stars}, and ignoring
possible interstellar reddening corrections\footnote{
Most of our stars are bright and have distances $<$1~kpc, so the reddening
should usually be small.
Moreover, we lack the information to assess reddening corrections accurately
for most of our targets.}
, we constructed the CMD shown in the top panel of Figure~\ref{fig4}.
Added to this figure is a shaded area that approximately outlines
the red horizontal branch CMD area identified empirically by \cite{kaempf05}
in $M_V$ versus $B-V$ coordinates.
In this figure panel the majority of our stars have CMD
locations consistent with RC/RHB membership, and there is no
apparent segregation between Li-rich and Li-poor stars.
On average they occupy the same evolutionary states.

In the bottom panel of Figure~\ref{fig4} we plot \teff\ and
\logg\ values of the program stars estimated from $Gaia$ EDR3 \citep{GAIA21}
photometry.
Here the shaded area represents the RC/RHB \teff\ domain used by
\cite{afsar18a}.
There are fewer data points in this Figure~\ref{fig4} panel than in the
top one simply because many program stars do not have $Gaia$ T$_{phot}$ and
\logg$_{phot}$ values.
But inspection of both upper and lower panels suggests the same conclusion: 
most stars reside in the RC/RHB region, and on average the locations of 
Li-poor and Li-rich stars are the same.

We extracted model atmospheres from the ATLAS grid \citep{kurucz11,kurucz18}.
The model effective temperatures were computed in steps of 100~K in the range
4000~$\leq$~\teff~$\leq$~5600~K.
The model gravities were estimated from recent empirical temperature-gravity 
correlations for metal-rich giant stars: Figure~8 of \cite{afsar18a}, and 
Figure~3 of \cite{casey19}.
At the high temperature end nearly all stars are on the core helium-burning 
horizontal branch, so we adopted a uniform \logg~=~3.0 in the range 
5300$-$5600~K.
For cooler temperatures the adopted gravities were decreased steadily,
ending with \logg~=~1.3 at \teff~=~4000~K.
Model metallicities were assumed to be solar, [M/H]~=~0.0, and microturbulent
velocities were assumed to be \vmicro~=~2.0~\kmsec; neither of these 
model parameter choices were critical for our desired photospheric spectrum
computations.

Whenever possible, we used observed $V-J$ colors and the color-\teff\ 
relations of \cite{ramirez05} to select appropriate \teff\ values
for each star, adopting a model atmosphere with the parameters as 
described above.
When $J$ magnitudes were not available we used $B-V$ colors, and in a handful
of cases we adopted models near those derived in individual papers, or
simply guessed at a starting model parameter set.
In this procedure we ignored possible interstellar reddening because our stars
are mostly near enough to not suffer large amounts of extinction, as noted
above.
We re-emphasize that we used the modeling only to reasonably match
the photospheric spectrum transitions.
Our procedures were entirely inadequate for detailed photospheric analyses,
and so we do not publish the model parameters adopted here.

\subsection{Observed/Synthetic Spectrum Matches}\label{specmatch}              

\begin{figure}
\epsscale{0.70}
\includegraphics[scale=0.70,angle=-90]{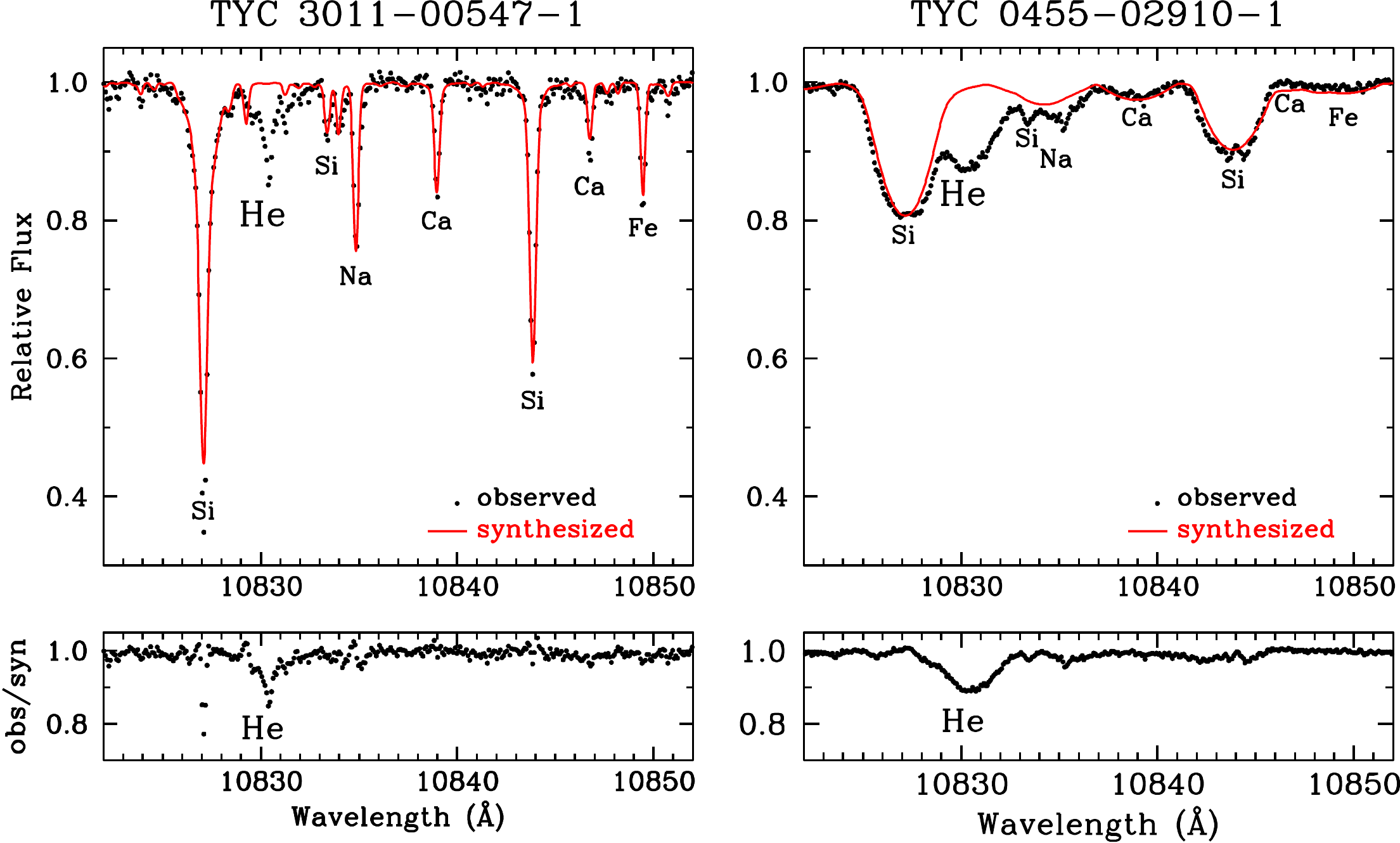}
\caption{\label{fig5}
\footnotesize
   Examples of photospheric spectrum cancellation.
   For each star the top panel shows an observed spectrum (black points) and 
   the best-fit model photospheric spectrum (line), and the bottom
   panel shows the result of the observed/computed division.
}
\end{figure}

We computed synthetic spectra of the stars in the 10820$-$10960~\AA\
region (the complete HPF spectral order containing $\lambda$10830), using the 
line list described in \S\ref{linelist} and model atmospheres in \S\ref{modatm}.
The computed synthetic spectra were compared interactively with the observed
spectra. 
The syntheses included combined thermal, microturbulent, and collisional 
damping line broadening effects, while the observed lines included these
and additional broadening due to instrumental, macroturbulent, and in 
some instances rotational effects.
These extra broadening contributions were accounted for in most cases by
Gaussian smoothing functions with empirically-determined broadening 
values that minimized the differences between observed and synthetic
spectra.
From iterative spectrum matches we were able to determine Gaussian
full-width-half-maxima values to precisions of $\sim\pm$0.02~\AA.
For some stars the smoothing functions combined Gaussian and rotational
factors.
In \S\ref{smoothing} we discuss spectrum line broadening in more detail.

We then performed a rough abundance determination in order to best match
synthetic and observed spectra near 10830~\AA.
We began with setting the N abundance, because the CN (0$-$0) Q-branch
bandhead culminating at 10925~\AA\ and the (0$-$0) R-branch bandhead at
10870~\AA\ often dominate those spectral regions (see Figure~\ref{fig3}),
and weaker CN lines occur in the whole spectral order.
Then we altered the Fe, Si, and Ca abundances to minimize the
differences between the synthetic and observed atomic lines.
Although \species{Na}{i}, \species{Ti}{i}, and \species{Sr}{ii} have 
significant absorption lines in this spectral order, none of them are
close enough to the \species{He}{i} $\lambda$10830 transition to matter
for our work.

The final step was to divide stellar and synthetic photospheric spectra to 
yield an essentially pure chromospheric spectrum of the \species{He}{i} line.
In Figure~\ref{fig5} we show examples of the \species{He}{i} isolation and 
measurement process for a typical program star with a modest \species{He}{i} 
line, and for a rapid rotator with a much stronger line.
For the left-hand panels, TYC~3011-00547-1 was chosen to illustrate typical 
$S/N$ conditions and the limits to our ability to completely extinguish the 
photospheric spectra.
In the right-hand panels we show the same observed, synthetic, and resultant
spectra for the rapidly-rotating star TYC~0455-02910-1.
Contamination by neighboring photospheric features is the dominant uncertainty 
source here.
But the $\lambda$10830 line clearly is strong in the divided spectrum.

\subsection{Equivalent Widths}\label{ewmeasures}

We measured $EW$s of the $\lambda$10830 lines using the specialized 
spectrum analysis code SPECTRE \citep{fitzpatrick87}\footnote{
Available at http://www.as.utexas.edu/$\sim$chris/spectre.html}.
The \species{He}{i} transition is usually very broad, with extended wings 
sometimes covering more than 2~\AA. 
Its line profile often has small distortions due to incomplete
cancellation of telluric and photospheric contaminating features. 
Therefore we measured $EW_{\rm He}$ via direct Simpson's Rule\footnote{
\eg, see https://math24.net/simpsons-rule.html}
integrations.
The $EW_{\rm He}$ values are listed in Table~\ref{tab-stars}.

\species{He}{i} absorption always is present in our stellar sample, albeit
very weak and hard to measure in a few stars.
There is a very large star-to-star $\lambda$10830 equivalent width range, 
$EW_{\rm He}$~$\simeq$~20$-$1300~m\AA.
Therefore for the remainder of this paper we will quote reduced widths for 
the 10830~\AA\ lines:
\begin{equation}
{\rm log(}RW_{\rm He}) \equiv\ log_{10}(EW_{\rm He}/\lambda) = log_{10}(EW_{\rm He}/10830)
\end{equation}
where $EW_{\rm He}$ is in units of \AA.
The \rwhe\ range in our stars is $\simeq$ $-$5.7 to $-$3.6.

For TYC~3011-00547-1 (Figure~\ref{fig5} left panels) we measured 
\rwhe~=~$-$4.94~$\pm$~0.04 ($EW$~=~125~$\pm$~13~m\AA), with the 
large uncertainty value based mostly on continuum placement, which directly 
impacts the wavelength extent of the \species{He}{i} line wings.  
For TYC~0455-02910-1 (Figure~\ref{fig5} right panels) we found
\rwhe~=~$-$4.50~$\pm$~0.03 ($EW$~=~345~$\pm$~15~m\AA).
Our measurements of $\lambda$10830 in the full sample yielded an estimated
general \rwhe\ uncertainty of $\pm$0.04~dex, which we will adopt hereafter
in this paper.

\subsection{Line Broadening}\label{smoothing}

\begin{figure}
\epsscale{0.70}
\plotone{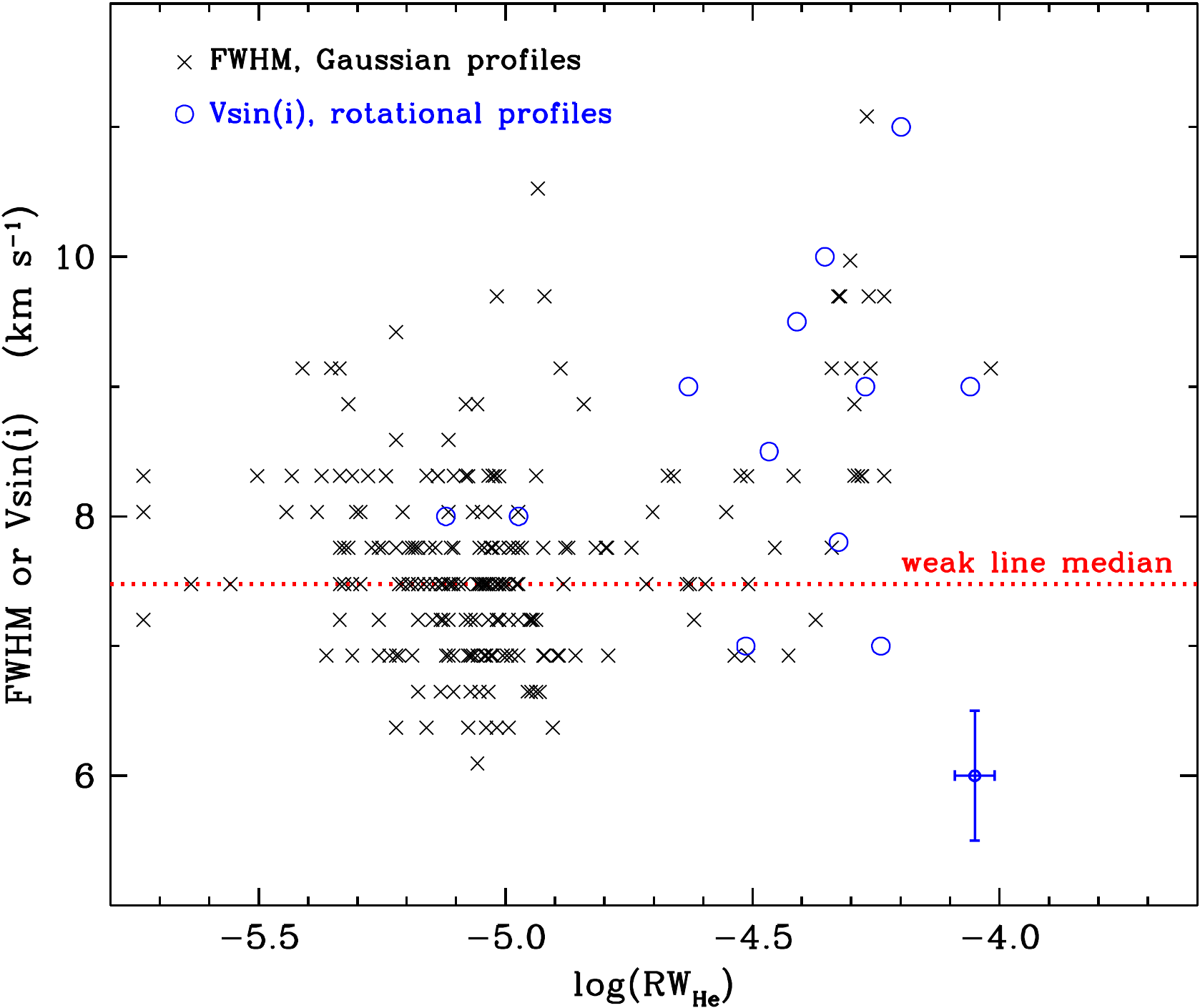}
\caption{\label{fig6}
\footnotesize
   Empirical broadening estimates compared to \species{He}{i} $\lambda$10830 
   reduced widths for stars that have relatively narrow line profiles.
   Point, line, and color types are identified in the figure.
   The ordinate of the plot is restricted to 0$-$12~\kmsec\ in order to see
   individual points clearly.
   This means that stars with \vsini~$\gtrsim$~12~\kmsec\ are not shown here.
   The point with error bars in the lower right-hand corner represents the 
   typical uncertainties of the measurements, as explained in the text.
   The $FWHM$ values were estimated to the nearest 0.01~\AA\ 
   ($\simeq$0.3~\kmsec), as can be seen in their ``quantized'' values.
}
\end{figure}

As described in \S\ref{specmatch}, for spectra without detectable rotational
line broadening we used comparisons of observed and synthetic spectra to 
estimate full widths at half-maximum ($FWHM$) values for the Gaussian smoothing
functions that account for instrumental and macroturbulent broadening.
In Figure~\ref{fig6} we correlate these $FWHM$ estimates with 
\species{He}{i} reduced widths.
The ordinate of this figure covers only the line breadth range
5~$<$~$FWHM$~$<$~12~\kmsec, allowing one to see that our $FWHM$ estimates
were quantized.
That is, in the synthesis/observation comparison the $FWHM$ values were
changed in steps of 0.01~\AA, which translates to the spacing of 
$\simeq$0.28~\kmsec\ that can be seen in Figure~\ref{fig6}.
For stars with quiet chromospheres, \rwhe~$\lesssim$~$-$4.8,
the observed photospheric-line widths appear to be constant to within our
measurement uncertainties.
Neglecting the few aberrant points with $FWHM$~$\gtrsim$~9~\kmsec, the mean 
value is $<$FWHM$>$~=~0.28~\AA\ ($\sigma$~=~0.03~\AA), or 
7.7~\kmsec\ ($\sigma$~=~0.5~\kmsec).
The median value is slightly lower: 7.5~\kmsec, which is 
indicated by the dotted red line in Figure~\ref{fig6}.

The instrumental resolving power $R$~=~55,000 contributes 5.5~\kmsec\ to 
the FWHM.  
We cannot determine directly a macroturbulent velocity, but the comparison 
between observed and instrumental $FWHM$ values suggests that 
V$_{mac}$~$\simeq$~2.2~\kmsec, a number that is nearly identical to our
assumed \vmicro~=~2.0~\kmsec.
If V$_{mac}$~$\sim$~\vmicro, then the empirically measured Gaussian 
smoothing values for stars with weak $\lambda$10830 lines appear to
successfully account for realistic line broadening in our red giants.
We adopt 0.5~\kmsec\ as a reasonable estimate of the $FWHM$ uncertainties
derived from our simple estimates.

In Figure~\ref{fig6} some stars are identified as rapidly rotating,
meaning that our synthetic-observed spectrum matches were improved by applying
a combination of rotational and Gaussian smoothing to the calculated syntheses.
Our rotational line broadening estimates were entirely empirical, based 
on standard relations\footnote{
\eg, http://www.astro.uvic.ca/$\sim$tatum/stellatm/atm6.pdf}, with an assumed 
limb darkening coefficient $u$~=~0.5 (see Figure~17.6 in \citealt{gray08}). 
We found through
repeated trials on many program star spectra that rotation could only be
detected with confidence at smoothing widths $\gtrsim$~10~\kmsec.
In the cases for which rotation could be detected, we set the Gaussian 
components (accounting for all broadening sources except rotation) to have
$FWHM$ values of 0.25~\kmsec.
The rest of the broadening was assumed to be caused by rotation.  
In Figure~\ref{fig6} the widths plotted for these stars are the
rotational \vsini\ values only.

There were some stars with derived $FWHM$~$\gtrsim$~10~\kmsec\ for which
we could not detect rotational broadening with certainty.
In these cases we have reported the Gaussian-only broadening.
It is likely that in some cases such stars have rotational signatures that
will be revealed with more rigorous analyses of spectra covering a much
larger wavelength range than that studied here.
The Gaussian or rotational broadening values for all stars are listed
in Table~\ref{tab-stars}.

\begin{figure}
\epsscale{0.70}
\plotone{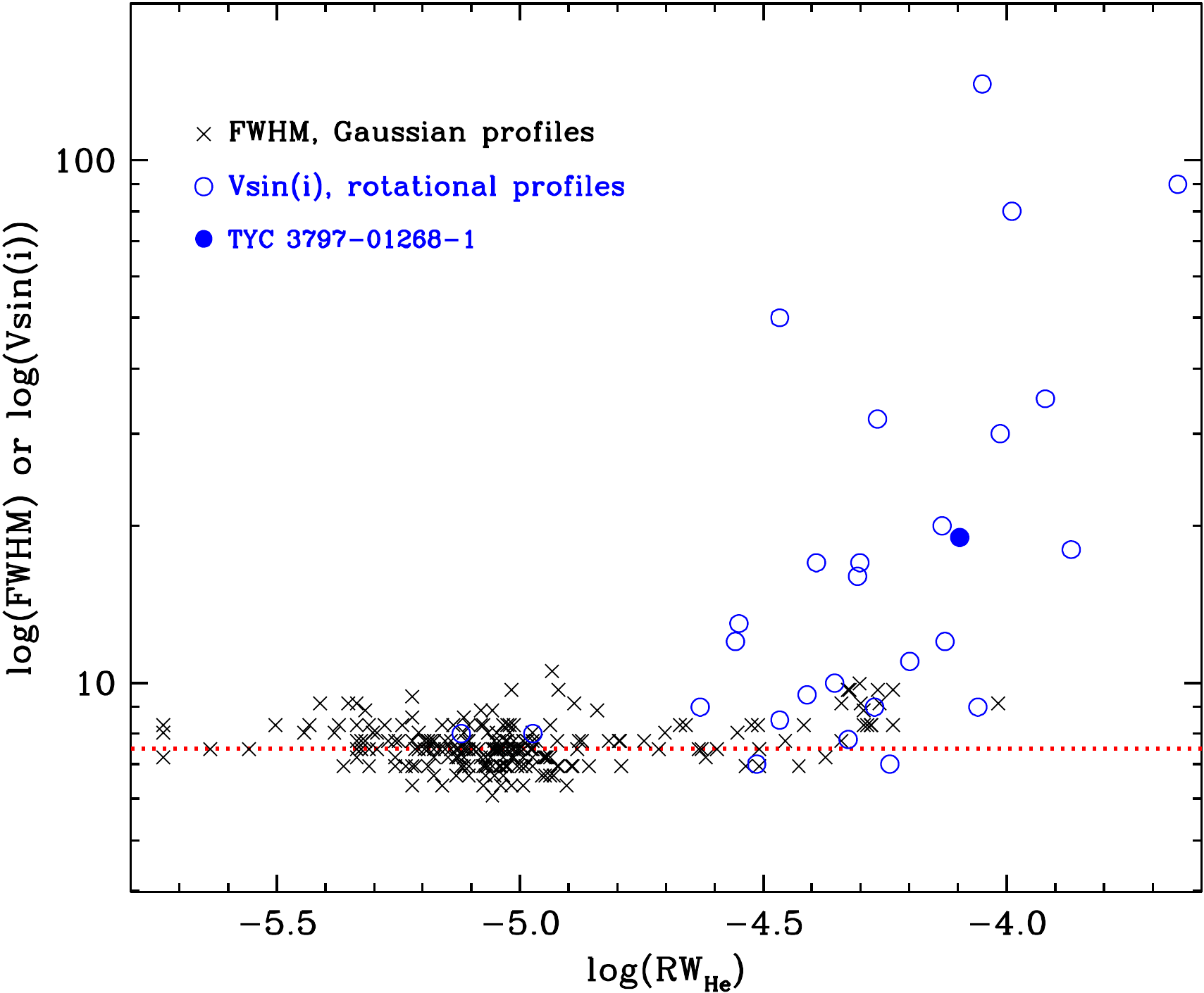}
\caption{\label{fig7}
\footnotesize
   Empirical broadening estimates compared to \species{He}{i} $\lambda$10830
   reduced widths for the whole sample.
   Point types and colors are as in Figure~\ref{fig6}, except that a filled
   circle is used to call attention of TYC 3797-01268-1, a program star
   with large infrared color excesses.
   This object will be discussed in \S\ref{lirichstars}.
}
\end{figure}

\begin{figure}
\epsscale{0.70}
\plotone{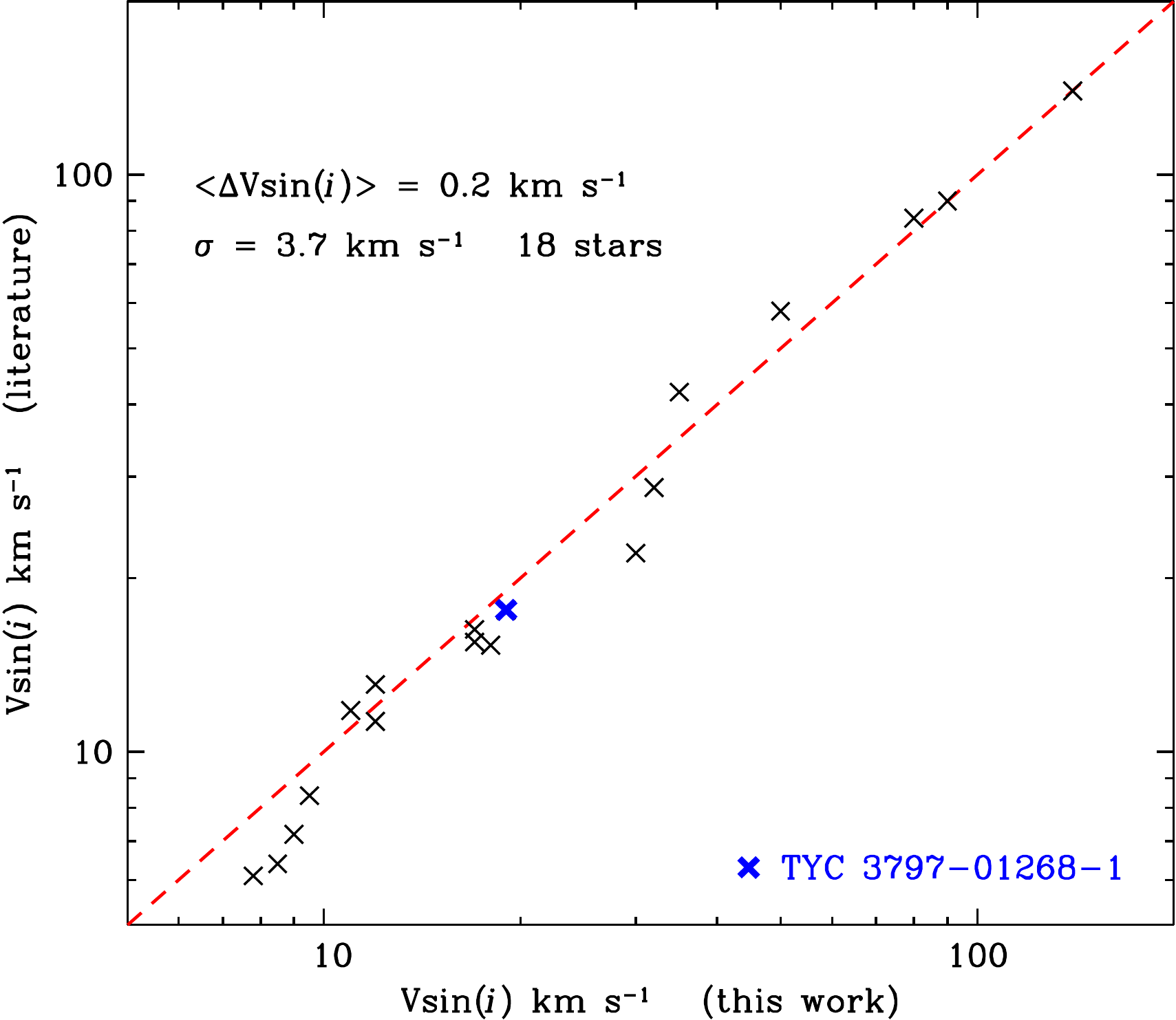}
\caption{\label{fig8}
\footnotesize
   Comparison of \vsini\ estimates in this work versus values published
   in various literature sources.
   The red dashed line indicates agreement between the two data sets.
   The thick blue $\times$ symbol marks TYC-3797-01268-1, noted earlier in
   Figure~\ref{fig7}.
}
\end{figure}

In Figure~\ref{fig7} we plot \rwhe\ versus $FWHM$ for all of the 
program stars.
Red giants with large $\lambda$10830 absorption features can have sharp
line profiles ($FWHM$~$<$~10~\kmsec) or can have obvious rotationally
broadened spectra.
But nearly all of our rapidly rotating program stars have very strong
\species{He}{i} absorption features (\rwhe~$\gtrsim$~$-$4.7).
There are only two exceptions in our sample.

We have searched the literature for published \vsini\ measurements of
our stars.
Various papers have used heterogeneous methods to derive \vsini\ from 
spectroscopic data sets with a variety of attributes.
In Table~\ref{tab-rotate} we list these literature values and their
sources.
Figure~\ref{fig8} shows a comparison of our rotational velocity estimates
with the published ones, and from the statistics quoted in the figure
legend, it is clear that they are in good agreement.
Note the small systematic offset at the low \vsini\ end of the correlation.
This indicates that translation of our measured line widths into \vsini\
values is probably too simplistic for the complex combination of instrumental,
thermal, microturbulent, macroturbulent, and rotational line components
at small rotational velocities.

\vspace*{0.2in}
\section{LI AND HE IN LI-POOR AND LI-RICH STARS}\label{results}

\begin{figure}
\epsscale{0.80}
\plotone{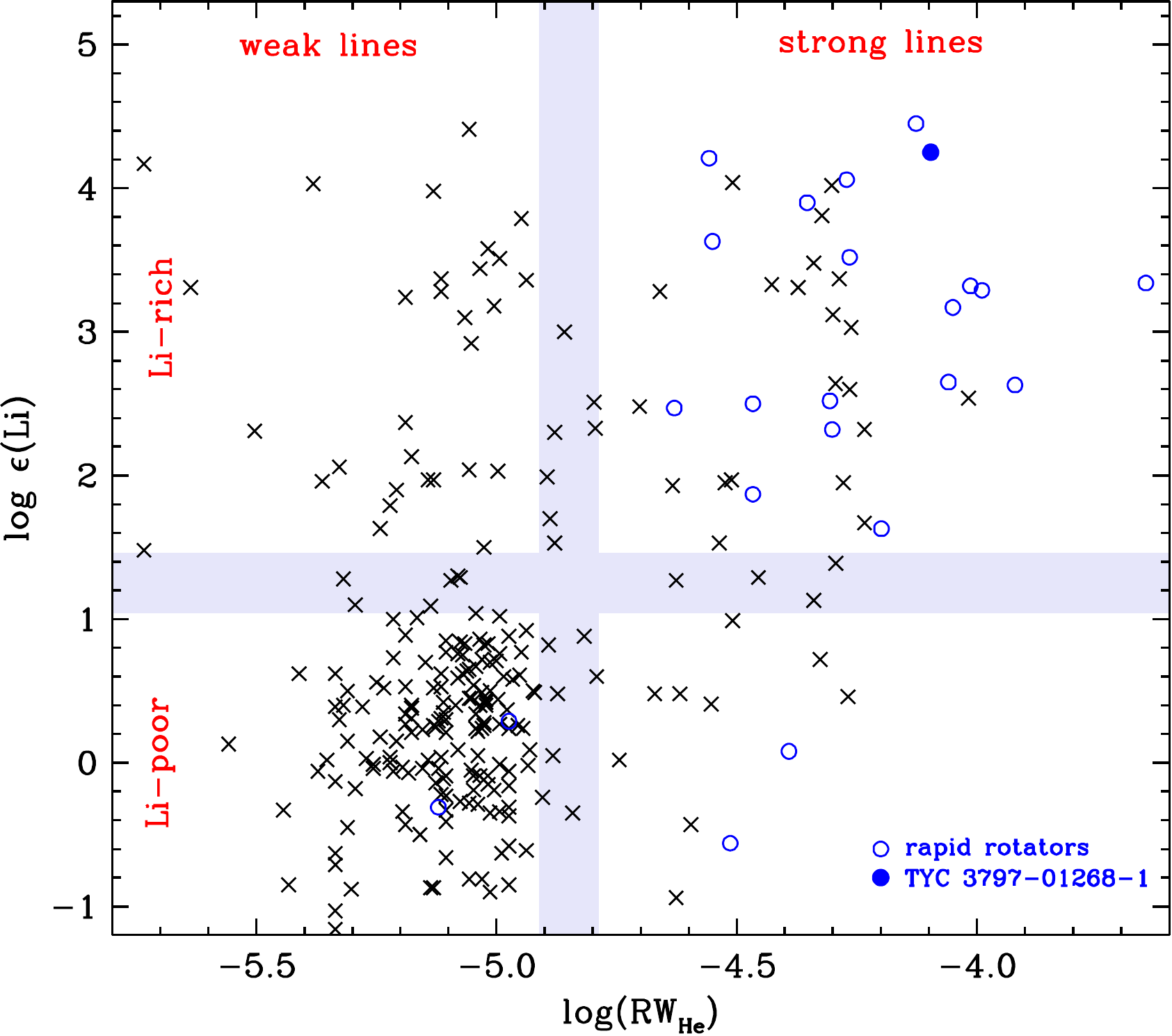}
\caption{\label{fig9}
\footnotesize
   \rwhe\ values plotted versus lithium abundances \eps{Li}.
   The horizontal shaded region serves to roughly divide Li-rich 
   from Li-poor stars, and the vertical shaded region separates the weak 
   and strong $\lambda$10830 absorption strengths.
   The location of these lines is discussed in the text.  
   As in Figures~\ref{fig6} and \ref{fig7}, black $\times$
   symbols represent stars with Gaussian measured line profiles,
   blue open circles are stars with rotational line profiles, and a blue
   filled circle denotes TYC 3797-01268-1 (see \S\ref{lirichstars}).
}
\end{figure}

\begin{figure}
\epsscale{0.80}
\plotone{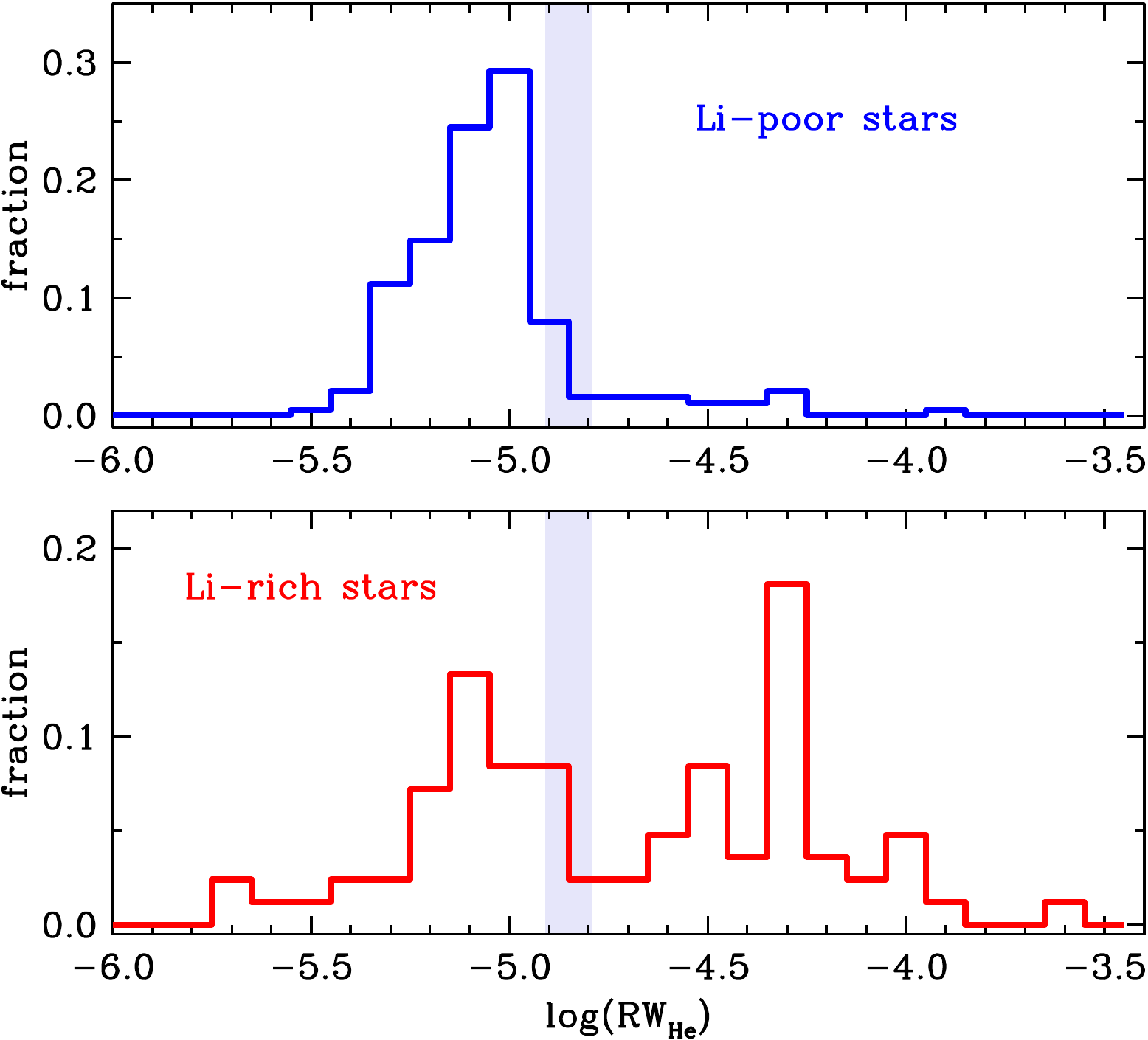}
\caption{\label{fig10}
\footnotesize
   Histograms representing the distribution of log($RW_{He}$) values Li-poor
   (upper panel) and Li-rich (lower panel) stars.
   The vertical shaded region, drawn at log($RW_{He}$) = $-$4.85 as in
   Figure~\ref{fig9}, separates the weak and strong $\lambda$10830 
   absorption strengths.
}
\end{figure}

Our basic result is that strong \species{He}{i} $\lambda$10830 absorption
is rare in Li-poor stars but very common in Li-rich ones.
This conclusion is clear in Figure~\ref{fig9}, where we compare Li abundances 
and \rwhe\ values, and in Figure~\ref{fig10}, where histograms of \rwhe\ are
shown for the Li-poor and Li-rich groups.
The horizontal shaded area centered at \eps{Li}~=~1.25 divides Li-poor 
and Li-rich abundance regimes, and the vertical band at \rwhe~=~$-$4.85 
divides weak and strong \species{He}{i} absorptions.
These lines create four apparent quadrants of different Li abundances and He
strengths, as labeled in the figure.
These group separations were determined empirically, and they merit further 
discussion.

In this exercise one should keep in mind that our stellar sample is not
statistically robust.
First and foremost, all of our Li abundances are literature values, which
guarantees their heterogeneity.
We have made no attempt to put them on a common abundance system with
consistent atmospheric parameters.
This is of no consequence for stars with very large Li abundances.
However for stars with lesser abundance values, 
1.0~$\lesssim$~\eps{Li}~$\lesssim$~1.5, assignments into Li-rich or 
Li-poor categories should be viewed with caution.
Secondly, our Li-rich stars were chosen through a literature search for 
reports of such stars.
Some of the papers considered the relative frequency of Li-rich stars, and 
others concentrated mostly on the properties of individual Li-rich objects.
Our Li-rich stars have irregular literature histories.

\vspace*{0.2in}                                               
\subsection{Li-poor Stars with Weak \species{He}{i} $\lambda$10830 Absorptions}\label{lipoorheweak}

For the Li-poor stars we used mostly candidates from the surveys of 
\cite{adamow14} and \cite{afsar18a}, neither of which were biased in any 
way on Li abundances.
The \citeauthor{adamow14} sample was the set of red giants of the 
Penn State-Toru{\' n} Planet Search (PTPS) program (\citealt{deka18} and 
references therein).
Their Li abundances were derived in that paper and thus took no part in 
sample selection.  
The \citeauthor{afsar18a} target list was formed from CMD searches for stars
that appeared to be in the He-burning red horizontal-branch domain, ignoring
possible Li abundance information. 
The Li abundances for these stars in Table~\ref{tab-stars} will be presented
by Bozkurt \etal\ (in preparation).
In general, our Li-poor stars have an abundance spread that is typical for
normal red giants.

The observed distribution of \rwhe\ among Li-poor stars in the 
lower left quadrant of Figure~\ref{fig9} shows a sharp drop in number
near \rwhe~$\simeq$~$-$4.85, with an uncertainty of about $\pm$0.05
by visual inspection.
We have 187 Li-poor stars as defined earlier (\eps{Li}~$<$~1.5).
Of these, 168 stars have \rwhe~$<$~$-$4.85, or 90\%, while only
18 (10\%) have stronger $\lambda$10830 \species{He}{i} absorptions.
Small alterations of the Li abundance and He line strength dividing lines 
would do quantitatively little to this result.
The vast majority of red giants have small Li abundances and weak 
\species{He}{i} $\lambda$10830 transitions.
For the rest of this paper we adopt \rwhe~=~$-$4.85 as the estimated break
point between stars with weak and strong $\lambda$10830 transitions.

\begin{figure}                                                
\epsscale{0.85}                                               
\plotone{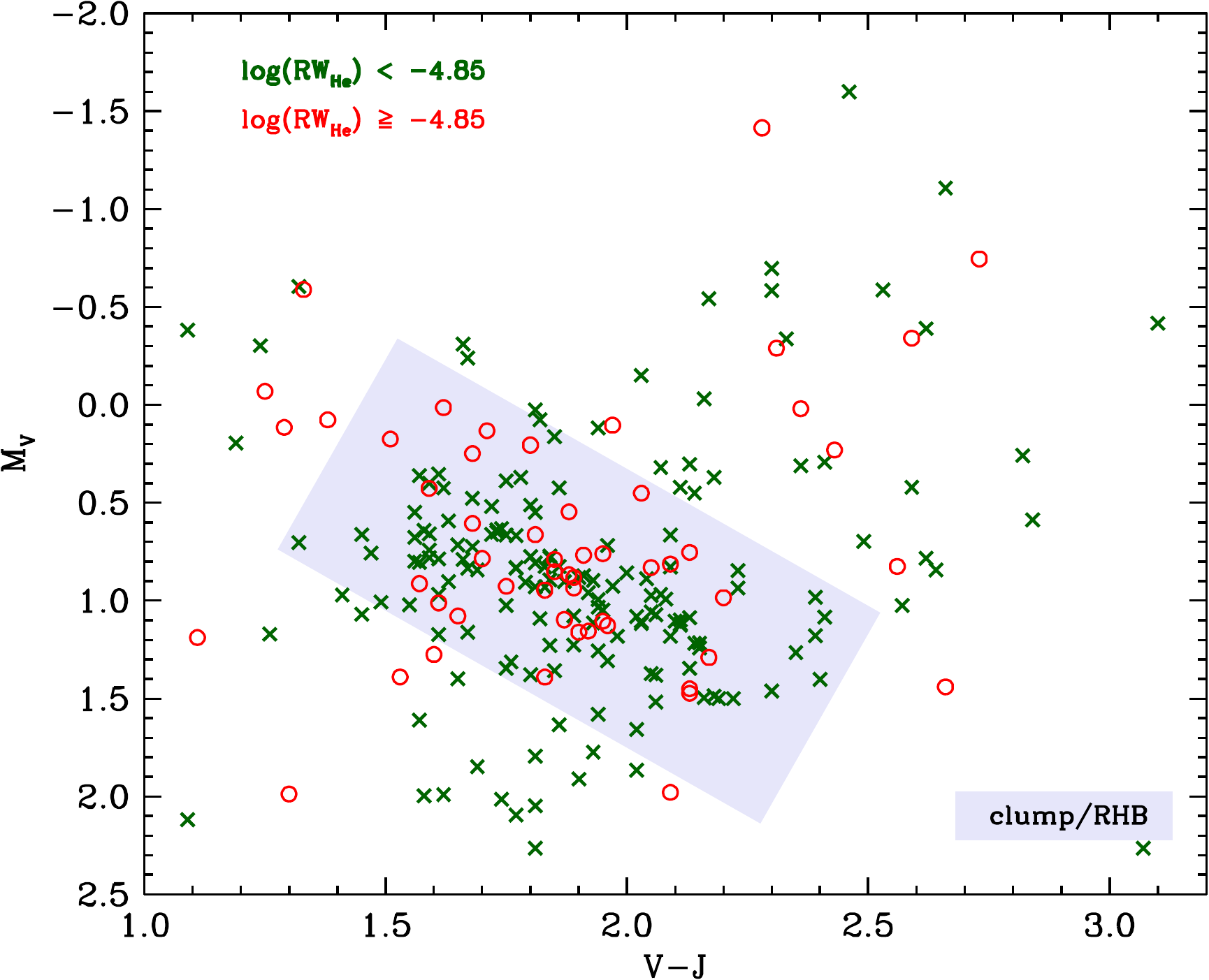}                                           
\caption{\label{fig11}                                        
\footnotesize                                                 
   Another CMD for the targets.
   This plot contains the same data that are displayed in the top panel of
   Figure~\ref{fig4}, but here the stars are divided into He-weak
   and He-strong groups, as indicated in the figure legend.
}                                                             
\end{figure}

With this \he\ estimated strength boundary we revisit the CMD of 
Figure~\ref{fig4}, and in Figure~\ref{fig11} we show the M$_{\rm V}$
versus V-J diagram with points separated at \rwhe~=~$-$4.85.  
Comparison of this figure to the top panel of Figure~\ref{fig4} reveals
general similarity.
The majority of both He-weak and He-strong stars reside in the
clump/RHB domain, and there is no obvious separation between the CMD 
distributions of the two He groups.  
Strong \he\ absorption features, like large Li abundances, appear most 
frequently in red giant clump and red horizontal branch stars.

Investigation of He absorption in red giants began with the moderate 
resolution ($R$~$\sim$~14,000) image-tube survey by \cite{zirin82} of 455 
stars of interest identified from \species{Ca}{ii} K-line measurements, 
and X-ray, and variability studies.
\cite{obrien86} provided the pioneering high-resolution ($R$~$\simeq$~54,000),
high signal-to-noise (typically $\gtrsim$100) survey of $\lambda$10830 for
G-K-M giants and supergiants.  
Many of their targets were either luminous giants and supergiants 
(luminosity classes I-II) or very cool ones (temperature class M).
Such stars often have large and variable \species{He}{i} $\lambda$10830
profiles, usually dissimilar to those in the G-K giants of interest here.
However, \citeauthor{obrien86} identified a class of less luminous K giants, 
which they labeled ``$\beta$~Gem type'', that have relatively modest
$\lambda$10830 absorptions which do not appear to vary much with time.
With ($V-J$) and $M_V$ values constructed from SIMBAD \citep{wenger00} data,
we identified about 25 of their stars lying within the boundaries of 
Figure~\ref{fig4}.
A direct comparison of their $EW$ values with our measurements is not possible
because their sample has only very bright stars ($J$~$<$~3.0) that cannot
be observed with HET/HPF.
However, we estimated a median \rwhe~$\sim$~$-$4.95 for their less luminous
K giants, less than our estimated upper limit for weak $\lambda$10830 
lines.
The median for our Li-poor stars, \rwhe~$\sim$~$-$5.10, is somewhat smaller
than that from \cite{obrien86}, but with different instruments, different \rwhe\
measurement techniques, and lack of stars in common this small issue
will not be pursued further.

\vspace*{0.2in}
\subsection{The Li-poor/Li-rich Split}\label{lipoorrichsplit}

Papers discussing Li abundances in red giants usually adopt \eps{Li}~=~1.5
as the minimum abundance for a star to be considered as Li-rich, and we have
followed this practice earlier in the paper.
This value goes back to the \cite{iben67a} stellar evolutionary computations,
leading to ``a standard first dredge-up dilution factor of 
$\sim$60, for a star with an initial cosmic abundance of 3.3~dex''.
\citep{charbonnel20}.
However, today we recognize that
the actual distribution of Li in evolved stars is the result of a complex
set of processes involving stars of different masses beginning with a variety
of initial conditions, followed by Li destruction mechanisms throughout the 
main sequence and subgiant evolutionary phases.
Several recent papers consider this point, including \cite{aguileragomez16},
\cite{casey19}, \citeauthor{charbonnel20}, \cite{kumar20}, and \cite{martell21}.
It is evident from these studies that Li enhancement is mostly associated 
with RC giants in the He-core burning phase. 
\cite{singh21} proposed a re-classification of RC giants based on Li 
abundances and asteroseismic properties into three categories: Li-normal 
(\eps{Li}~$<$~1.0), Li-rich (1.0~$<$~\eps{Li}~$<$~3.2), and super-Li-rich 
(\eps{Li}~$>$~3.2). 
Thus the ``traditional'' Li-rich marker, \eps{Li}~=~1.5$-$1.8 depending on 
mass, is now recognized as convenient upper limit for giants ascending the 
RGB for the first time but too simplistic in describing the Li abundance 
history of real stars.  
The only reported Li-rich first-ascent RGB giant that has been classified using 
asteroseismic data is KIC~9821622 \citep{jofre15b}.
\cite{singh20} have re-analyzed this star and derived a lower Li abundance, 
substantially smaller than the suggested super-Li-rich category.
As of today we do not know of any giant with \eps{Li}~$>$~1.8 that is
classified as a giant ascending the RGB with a He-inert core based on
astroseismic analysis; certainly no super Li-rich RGB giant has been found
\citep{singh19b}. 

Our assessment of the Li-poor/Li-rich division comes simply from inspection
of the distribution of points in Figure~\ref{fig9}.
The most obvious separation of stars in Li abundance occurs near
\eps{Li}~$\simeq$~1.0, with Li-poor stars aggregating at small \rwhe\
values while Li-rich stars have much larger star-to-star \rwhe\ scatter.
Therefore, we have chosen to set the Li-rich/Li-poor split at 
\eps{Li}~=~1.25, with an uncertainty of at least 0.1~dex.
We emphasize again that this line is a convenient demarcation between the
two Li stellar groups, and it should not be considered a well-determined rigid
boundary.

\vspace*{0.2in}
\subsection{The Li-rich Red Giants}\label{lirichstars}

Our sample has 84 stars with \eps{Li}~$>$~1.25.
Of these, 46 have \rwhe~$>$~$-$4.85 (55\%).
This high fraction of Li-rich stars with strong $\lambda$10830 absorptions
can be easily seen in Figure~\ref{fig9} and in the bottom panel of
Figure~\ref{fig10}.
Small adjustments of the Li abundance and/or \species{He}{i} line strength 
divisions do not disturb the basic positive Li/He connection in Li-rich
stars.

Almost all of the strongest $\lambda$10830 features are detected in
rapidly-rotating stars.
Nine of the 10 stars with \rwhe~$>$~$-$4.2 ($EW$~$>$~680~m\AA) have 
rotational metal line profiles.
But many Li-rich stars with $-$4.85~$<$ \rwhe\ $<$~$-$4.20 exhibit no 
excess line broadening. 
Therefore, in line with conclusions of literature Li abundance studies, it is 
clear that rapid rotation is not a requirement for either Li abundance 
excess or for strong \he\ absorption.
It is more certain that rotation is linked with strong \he\ lines.
Our sample has 28 stars with detectable rotational line profiles, and only two 
of them have weak $\lambda$10830 absorptions, \rwhe~$<$~$-$4.85 
(Table~\ref{tab-stars}).

Large rotation and strong $\lambda$10830 absorption do not guarantee 
large Li: three such stars in Figure~\ref{fig9} have \eps{Li}~$\lesssim$~0.0.
But Li enhancement among giants is a transient phenomenon; very large Li 
abundances decline with time, as shown by \cite{singh21}.
The consequence of this is that there may be some Li-poor giants that still
have strong \species{He}{i} $\lambda$10830 transitions and high rotation
rates.

The $\lambda$10830 triplet has become a powerful diagnostic of 
physical conditions and dynamics in many stellar and exoplanetary objects.  
As discussed in \S\ref{analysis} the lower energy level of the transition 
lies at 19.8~eV and is metastable.
This mandates that reasonably high chromospheric temperatures are required 
to produce the line; it cannot be a photospheric transition.  
Direct evidence from the Sun demonstrated that the transition could be produced
as a result of photoionization by radiation occurring shorter than 504~\AA, 
followed by recombination into the lower metastable level of the transition.
The relationship between X-ray radiation and the strength of the helium 
line in the solar case (\citealt{harvey75}, \citealt{cranmer09}) inspired 
early studies of the transition in cool stars (\citealt{zarro86}, 
\citealt{obrien86}).  
These authors found a strong correlation between the $EW$ of the $\lambda$10830
line and the value of L$_{X}$/L$_{bol}$ for cool dwarfs and for G and K giant 
stars.  
Sparse data exist for G and K supergiants, but a correlation appears likely.

A direct measure of the photoionizing flux in dwarfs and giants 
suggested that the photoionization and recombination process dominates in 
giant stars. 
However, active dwarfs and subgiants appear not to exhibit a correlation    
between the photoionizing flux and helium line strength, presumably due to  
their higher chromospheric densities \citep{sanzforcada08}.
                                                              
\begin{figure} \epsscale{0.85}
\plotone{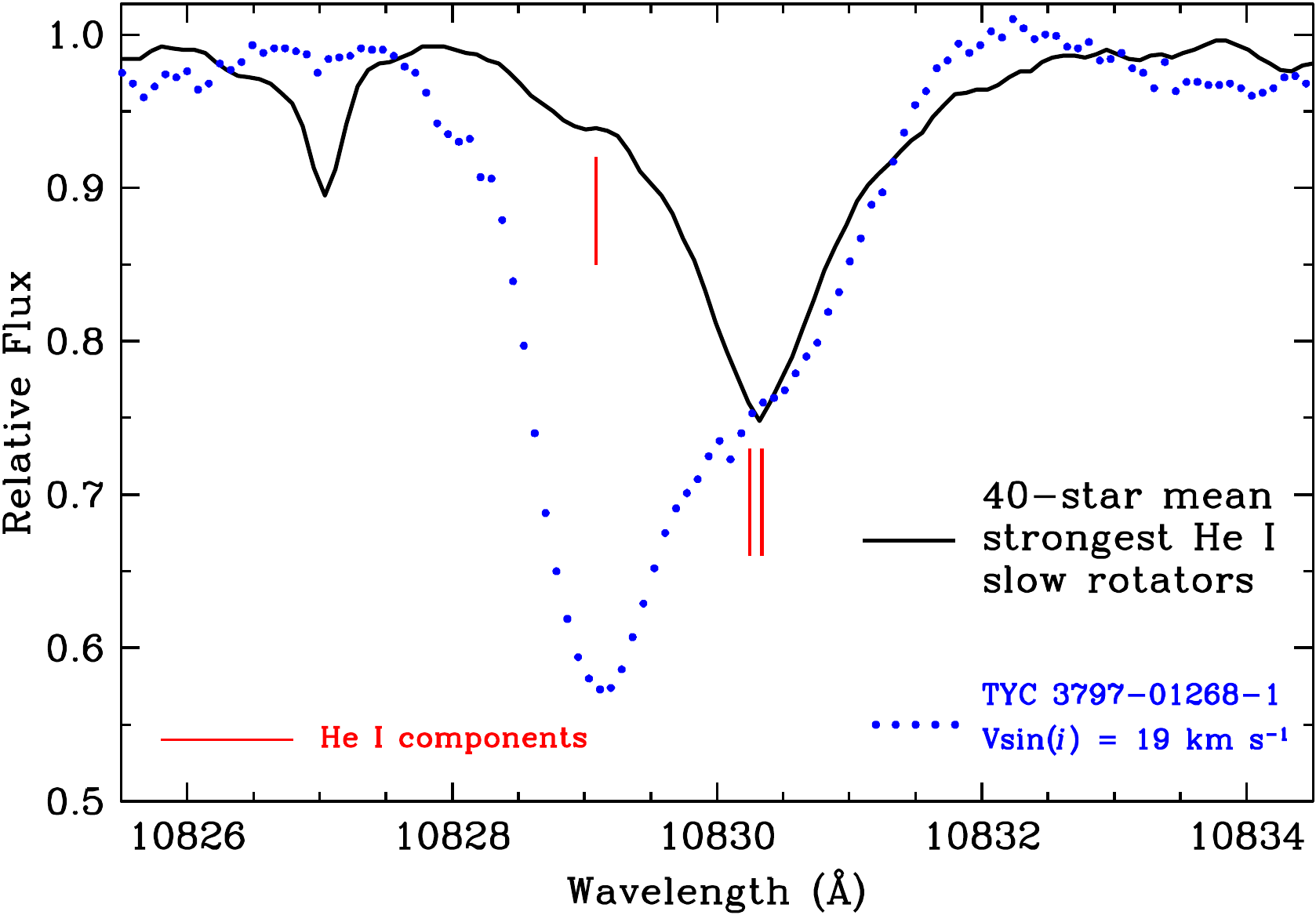}
\caption{\label{fig12}
\footnotesize
   A small section of spectra showing details of the He line profiles.
   The solid black line represents the mean line profile of the strongest 
   \he\ $\lambda$10830 transitions for program stars without significant 
   rotational broadening.
   The $\lambda$10830 line is a triplet, whose components are marked with
   vertical red lines.
   The feature at 10827~\AA\ is a residual of imperfect cancellation of
   the very strong photospheric \species{Si}{I} line, and should be 
   ignored.
   The dotted blue line shows our single observation of TYC~3797-01288-1.
   In the figure legend we quote our derived \vsini\ value.
   See the text for discussion of this unusual star.
}
\end{figure}

Another feature of the $\lambda$10830 transition derives from its 
metastable nature which makes the line profile a tracer of a stellar wind flow.
\cite{dupree96} detected blue-wing asymmetries of this transition, leading to
the inference of radial outflows from the Sun.                
This feature has also been seen in luminous stars (\citealt{obrien86},      
\citealt{obrien86}) including metal-poor field giants         
(\citealt{dupree92,dupree96}).                                
However, stellar winds are unlikely to be major contributors to the         
\he\ line formation in almost all of our stars.                            
In Figure~\ref{fig12} we show a mean line profile for the 40 stars with    
strongest $\lambda$10830 absorptions (\rwhe~$>$~$-$4.85) that either are 
without detectable rotation or are not rotating rapidly 
(\vsini~~$<$~10~\kmsec).    
The co-added spectra are those have already undergone photospheric spectrum
cancellation (\S\ref{specmatch}, \S\ref{ewmeasures}).         
Inspection of the mean profile reveals the presence of the weakest of       
the $\lambda$10830 triplet members at 10829.09~\AA, but otherwise it appears  
to have symmetric blue and red wings centered on the lab wavelength of the  
two main triplet members.                                     
There is no evidence for the blue asymmetries that are seen in some luminous 
stars of the above-cited papers.

\vspace*{0.2in}
\subsection{\he\ $\lambda$10830 in the Unusual Li-rich Star TYC 3797-01268-1}\label{tyc3797}

TYC 3797-01268-1 (HD~233517) has unique properties among our program
stars.
\cite{fekel96} first reported on the high lithium and rotation of this star.
In Tables~\ref{tab-stars} and \ref{tab-rotate} we quote the 
\cite{balachandran00} values of \eps{Li}~=~4.2 and \vsini~=~17.6~\kmsec.
\cite{jorissen20} used radial velocity and astrometric data to make a 
convincing case for a binary companion to TYC 3797-01268-1.

Our \he\ $\lambda$10830 line spectrum for this star is very strong:  
\rwhe~=~$-$4.10 ($EW$~=~867~m\AA).
But the line profile is very unusual for our program stars.
In Figure~\ref{fig12} we display the TYC 3797-01268-1 spectrum, and its
contrast to the 40-star mean of other strong $\lambda$10830 lines is 
striking.
The blue-wing asymmetry discussed in \S\ref{lirichstars} is a prominent
spectral feature of this star.
To signal this star's uniqueness we already have labeled its position in 
Figures~\ref{fig7}, \ref{fig8}, and \ref{fig9}.

\begin{figure}
\epsscale{0.85}
\plotone{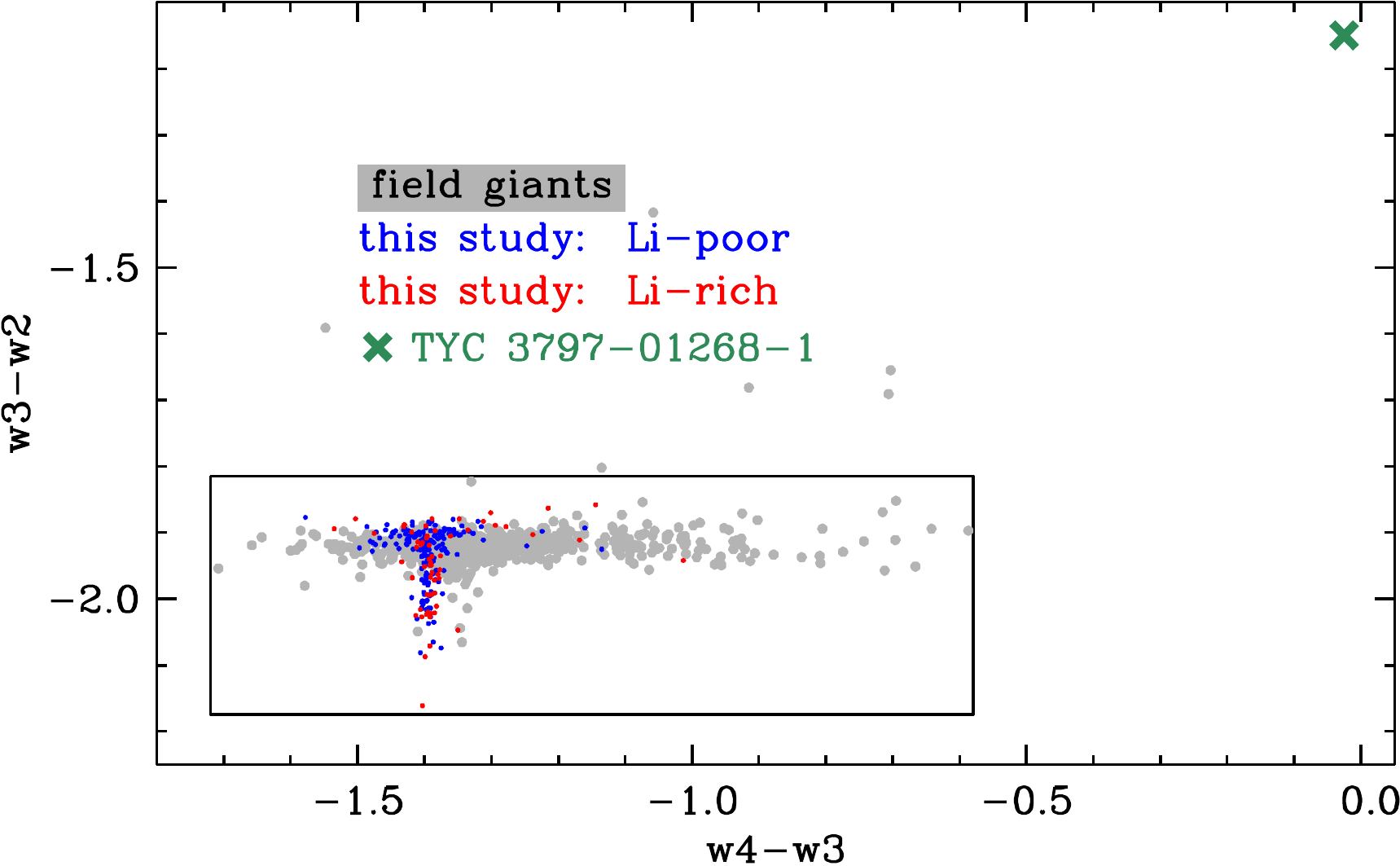}
\caption{\label{fig13}
\footnotesize
   A far-IR color-color plot from WISE.
   The long rectangular box defines the limits of these colors for  red 
   giant stars without infrared excesses, as computed by \cite{mallick22}.
   Special attention is called TYC~3797-01268-1 with a dark green $\times$
   symbol.  
   See \S\ref{tyc3797} for discussion of this star.
}
\end{figure}

Giants with very high lithium abundances and fast rotation can be 
understood if the Li enhancements have been caused by external events such as 
mergers with massive planets, brown dwarfs or compact objects like 
white dwarfs. 
An expectation for the merger scenario for Li-rich giants is the presence
of dust as a result of mass loss, and hence the infrared excess 
\citep{denissenkov04,carlberg10,zhang13}.  
However, IR excess among giants is extremely rare as shown by various 
surveys, \eg, \citet{zuckerman95,kumar15,mallick22}.
We searched for IR excess among our program stars by using Wide-field 
Infrared Survey Explorer (WISE; \citealt{wright10}) colors, as detailed in 
\citet{mallick22}. 
Of the 278 giants in our study 262 stars have flux measurements in all 
four WISE bands (3.3~$\mu$, 4.6~$\mu$, 11.6~$\mu$, 22.1~$\mu$m).
Figure~\ref{fig13} displays a WISE color-diagram for red giants in general and
for our program stars.
This plot is similar to the upper panels of Figure~4 in \citeauthor{mallick22}.
In particular, the rectangular box is the theoretical ``zero IR excess box''
computed by them based on the effective temperatures of  stars in the red 
giant/clump temperature domain.  
Observed colors with larger values of w4-w3 or w3-w2 than the box limits
are presumed to be affected by circumstellar disk material.

Inspection of Figure~\ref{fig13} reveals IR excess only in 
TYC~3797-01268-1.
This is in agreement with the \citeauthor{mallick22} conclusion that Li-rich 
giants with IR excess are more likely to be fast rotators.
Unfortunately, none of their giants are in our current program because 
their analysis is based on the GALAH spectroscopic survey in the southern 
hemisphere, largely inaccessible to HET/HPF observations.  
To explore this idea more completely, it will be necessary to analyze more 
rapid rotators among Li-rich giants that have strong $\lambda$10830 
transitions and IR excess. 
Our stellar sample, constructed with primary attention to Li abundance, does
not fairly represent the rapid-rotation red giant sub-population. 
A better survey based on a large \vsini\ range is appropriate for a
future study.

We also constructed flux curves for TYC~3797-01268-1
using broad-band colors from optical to far infrared wavelengths 
(0.4$-$100~$\mu$m) using the spectral energy distribution software package 
VOSA \citep{bayo08,bayo20}\footnote{
Available at http://svo.cab.inta-csic.es}.  
Comparison of the observed flux curves to theoretical ones confirms that
$IR$ excesses become apparent at wavelengths beyond $\sim$5~$\mu$m in this
star, a clean signature of a very dusty red giant.

Finally, we also examined individual $\lambda$10830 profiles in our 
sample of rapidly rotating stars.
These profile shapes sometimes are uncertain due to significant contamination 
by neighboring photospheric features, as seen in the right panel of
Figure~\ref{fig5}.  
However, almost all of these \species{He}{i} profiles appear to be reasonably
symmetric and centered at 10830.3~\AA.
The single case of a significant blue-shifted line profile is TYC~3797-01268-1.

\vspace*{0.2in}                                               
\section{STELLAR MOTIONS}\label{move}

\vspace*{0.2in}                                               
\subsection{Space Velocities}\label{spacevel} 
                                                              
\begin{figure}
\epsscale{0.85}
\plotone{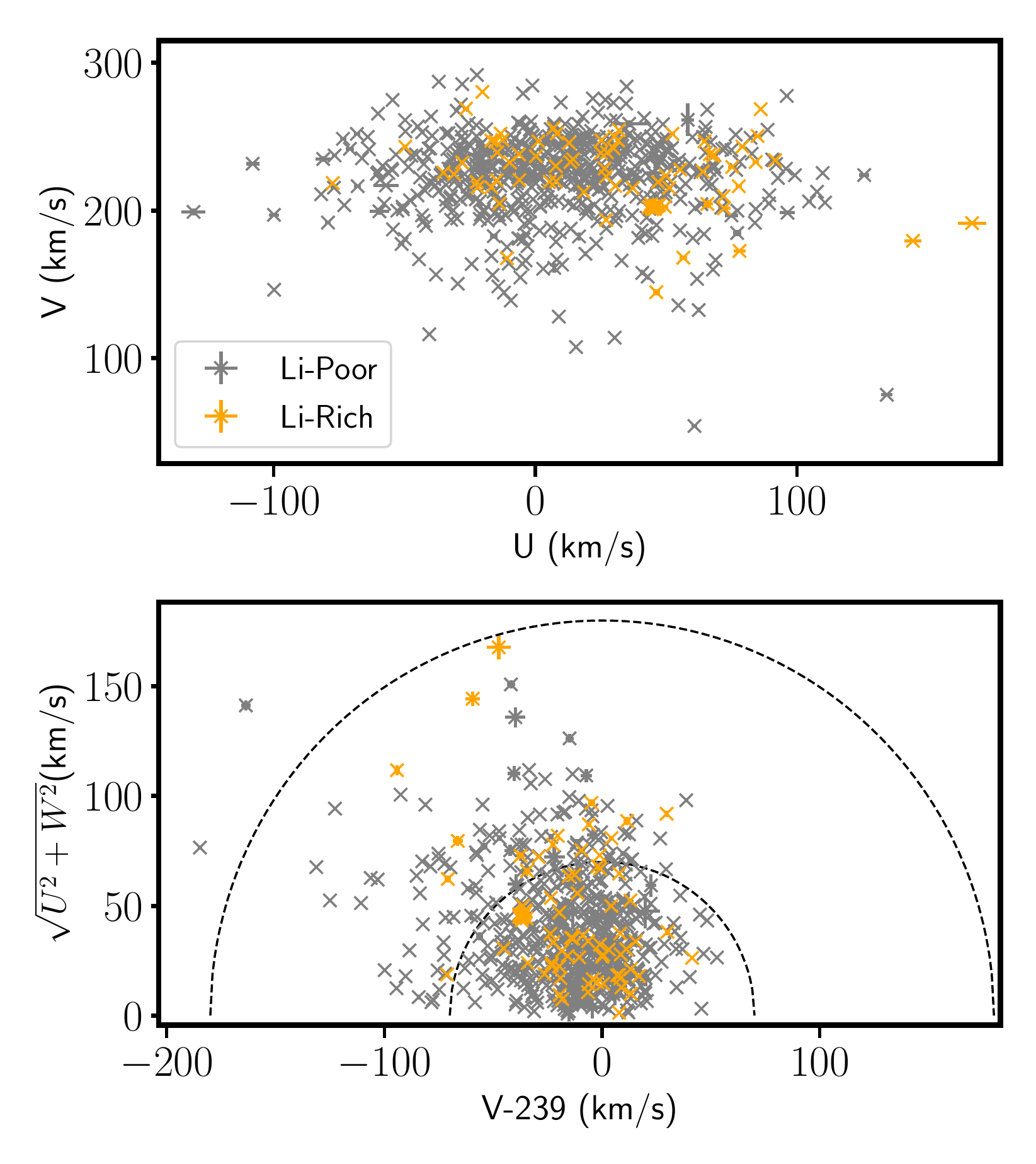}
\caption{\label{fig14}
\footnotesize{
   Top panel: U-V velocity plane for Li-rich stars (orange `x's) and 
   Li-poor stars (gray `x's). 
   Bottom Panel: Toomre Diagram  for Li-rich stars and Li-poor stars. 
   In the Toomre diagram panel, we subtract 239~\kmsec\ to place the 
   velocities in the local standard of rest. 
   For reference, the dashed lines represent a constant total velocity of 
   70 and 180~\kmsec, respectively. 
   No significant kinematic difference is detected between the Li-rich and 
   Li-poor populations.}
}
\end{figure}

In order to determine whether Li-rich and Li-poor stars belong to different 
kinematic structures, we obtain 3 dimensional (3-D) Cartesian velocities 
from \cite{marchetti21}. 
For reference, we define a 3-D Cartesian velocity (U, V, W) and position 
(X, Y, Z) vector for each star whereby U is positive in the direction 
pointing toward the GC, V is positive along the direction of the disk 
rotation, and W is positive when pointing towards the North Galactic pole. 
In this convention, we assume that the Sun's location is 
(X, Y, Z) = (8.20, 0.00, 0.025)~kpc and 
(U, V, W) = (14.0, 250.24, 7.25)~\kmsec\ relative to the Galactic 
center \citep[e.g.,][]{schonrich12}. 
The spatial positions and velocities for the stars in this study are 
computed using astrometric information from the early\footnote{
We note here that while the full third data release of Gaia was released 
in June 2022, the astrometric information was not updated between the early 
data release 3 and the full data release 3.} 
data release 3 from the Gaia spacecraft \citep[e.g., parallax, proper 
motion, and sky positions][]{GAIA21}. 
The radial velocities are also sourced from Gaia EDR3 \citep{GAIARV2019}. 
The astrometric and radial velocity information are then used to derive full 
3-D positions and velocities using the method outlined in section~2 
of \cite{marchetti21}. 
With the 3-D velocities for each star, we computed the probability 
that the star belonged to the thin disk, thick disk, or halo population 
using the method outline in section 2.4 of \cite{ramirez13}. 

In Figure~\ref{fig14}, we show the U-V kinematic plane (top panel) 
and the  Toomre Diagram (bottom panel) for the Li-rich stars (orange `x's)
and Li-poor/normal stars (gray `x's). 
From these plots we find that the Li-rich stars and Li-poor (normal) stars 
are mostly part of the kinematic thin disk and the two samples do not seem 
significantly different. 
To quantify this further, we also compute the fractions of Li-rich and Li-poor 
stars that are parts of the kinematic thick and thin disks using the 
probabilities defined above. 
We find that more than 80\% of the Li-rich stars have a 70\% or higher 
chance of belonging to the kinematic thin disk. 
Similarly, $\sim$90\% of the Li-normal/poor stars have a 70\% or higher 
chance of belonging to the kinematic thin disk. 
From this probabilistic kinematic analysis, as well as Figure~\ref{fig14}, 
we can conclude that the sample of Li-rich and Li-poor stars are almost 
entirely consistent (at the 80-90\% level) of thin disk giant stars. 
We performed a 2-sample Kolmogorov–Smirnov test, finding a p-value of 
0.43, on the total velocity distribution for the Li-rich and Li-poor stars. 
This test allows us to assess the probability that the Li-rich and Li-poor 
stars are drawn from the same underlying kinematic distribution. 
The results of this test indicate that there is not sufficient evidence 
that the Li-rich and Li-normal stars are drawn from different underling 
kinematic populations.
We also find that $\sim$90\% of the He strong (\rwhe~$>$~$-$4.85), have a 70\% 
or higher chance of belonging to the kinematic thin disk.

\vspace*{0.2in}
\subsection{Astrometric Accelerations with $Hipparcos$ and $Gaia$}\label{binarity}

\begin{figure}
\epsscale{0.85}
\plotone{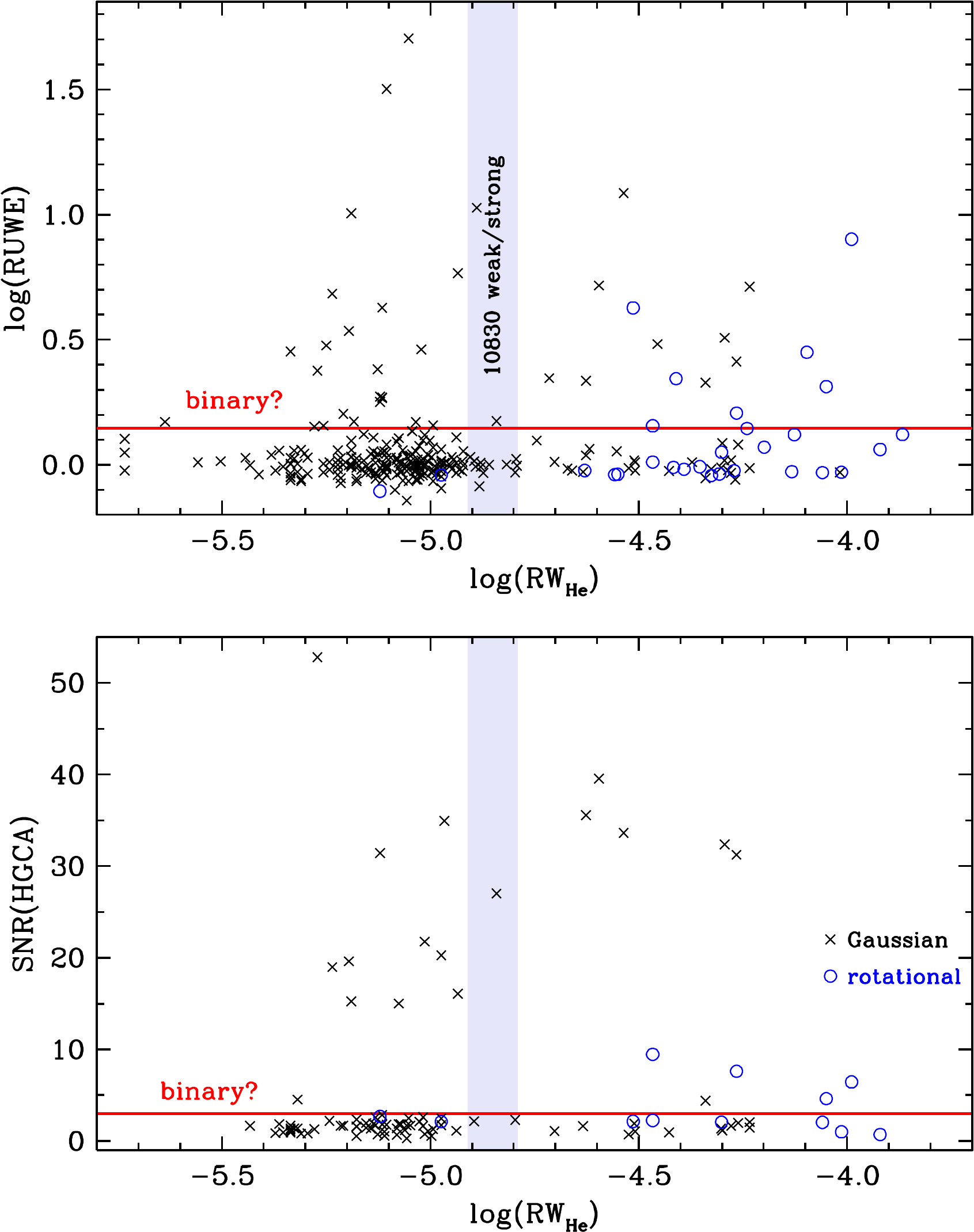}
\caption{\label{fig15}
\footnotesize
   Correlations of \rwhe\ with the $Gaia$ Renormalized Unit Weight 
   Error (RUWE; top panel) and $Hipparcos$-$Gaia$ Catalog of Accelerations 
   (HGCA; bottom panel).                                              
   The symbols are the same as in Figure~\ref{fig9} and noted in the       
   figure legend.                                             
   The red line in each panel indicates the minimum value of each of these  
   excess astrometric motion indicators that might suggest the presence     
   of binary motion.                                          
   See text for more details.
}
\end{figure}

\begin{figure}
\epsscale{0.90}
\plotone{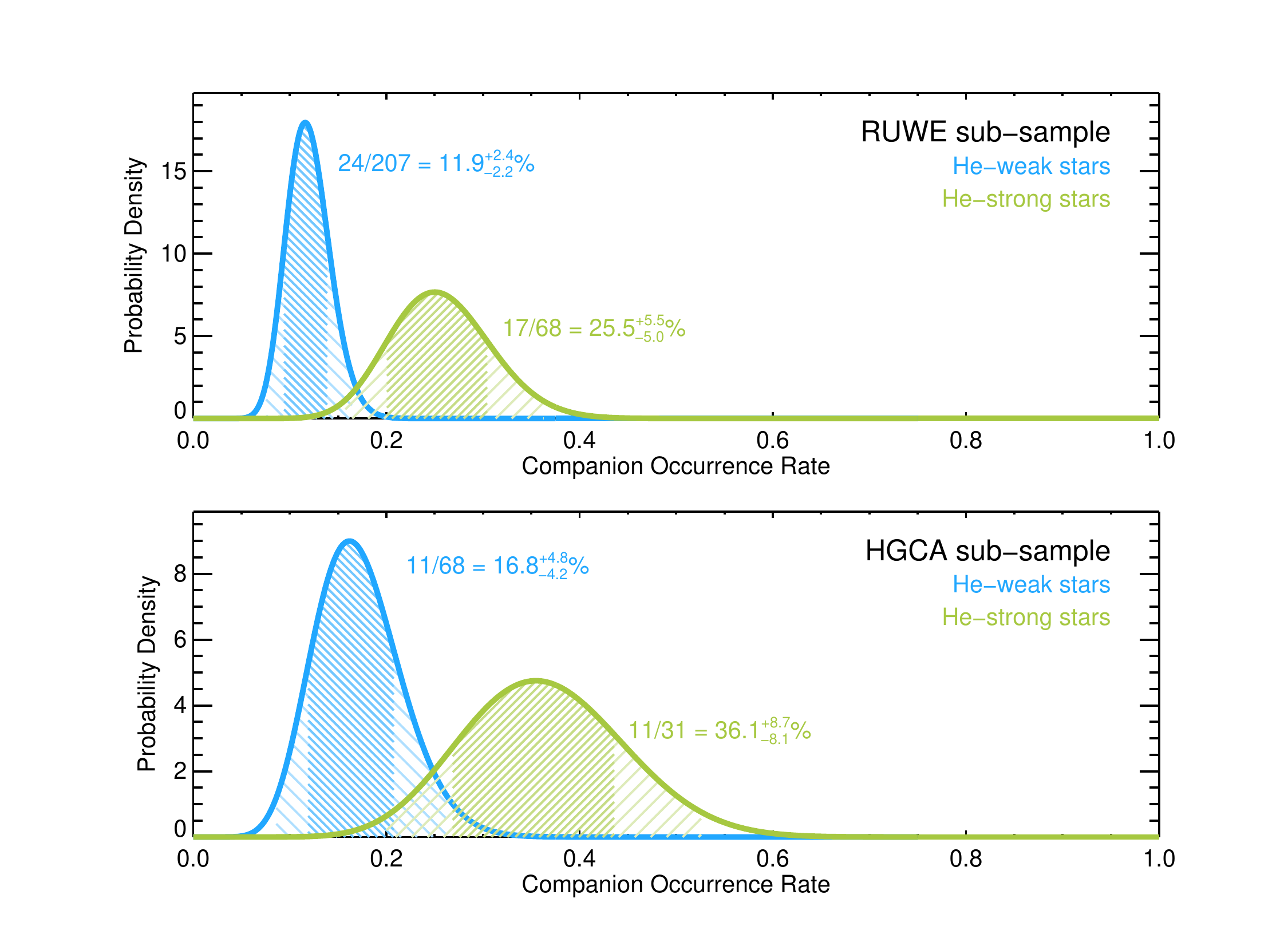}
\caption{\label{fig16}
\footnotesize
   Constraints on the frequency of binary companions for stars in our 
   sample with $Gaia$ RUWE measurements (top) and those with HGCA 
   accelerations (bottom).  
   In both cases, stars with strong helium absorption have a higher companion 
   occurrence rate at the 2.3-$\sigma$ level for the RUWE sub-sample and 
   at the 2.0-$\sigma$ level for the HGCA sub-sample.
}
\end{figure}

In the \cite{adamow14} spectroscopic survey, Li was detected in 82 
giants, with 11 of them being Li-rich ($A$(Li)$>$1.4).  
Among these, four showed evidence of stellar or substellar companions based 
on radial velocity variations.  
They speculated that excess lithium and binarity may be related, although 
the underlying basis and statistical validation of any such connection is 
not yet established.
More recently, \cite{goncalves20} and \cite{jorissen20} have conducted
radial velocity surveys of individual Li-rich stars, with mixed conclusions.
\citeauthor{jorissen20} found no excess binary frequency among their 11 Li-rich 
giants compared to Li-poor samples.
\cite{goncalves20} studied 18 Li-rich stars and detected radial velocity
variations in some of them, but suggest that further velocity monitoring work
is needed before reaching final conclusions on binarity statistics.
Systematic radial velocity campaigns to increase sample sizes would be welcome.

To help assess the binary frequency in our Li-rich stars, we have 
cross-matched the 278 stars in our program with $Gaia$ EDR3 \citep{GAIA21} 
and the $Hipparcos$-$Gaia$ Catalog of Accelerations (HGCA; \citealt{brandt18}; 
\citealt{brandt21}).  
$Gaia$'s Renormalized Unit Weight Error (RUWE) value provides information about
the goodness-of-fit of the nominal five-parameter astrometric solution.
Values near 1.0 are expected for a typical good fit, whereas values larger 
than $\approx$1.4 have been shown to indicate the presence of a stellar 
companion \citep{stassun21}.
HGCA makes use of recalibrated proper motions from $Hipparcos$ and $Gaia$ as 
well as the mean scaled positional difference between the two epochs to 
measure long-term accelerations (changes in proper motion) over this time 
period.  
The most precise acceleration measurement that we make use of in this study is 
between $Gaia$ and the $Hipparcos$-$Gaia$ scaled positional difference.
The 34-month baseline sampled in $Gaia$'s EDR3 release is most sensitive to 
companions at separations of a few AU, whereas the $\approx$25-year baseline 
of the HGCA catalog can reveal companions at typical separations of a few to 
tens of AU (e.g., \citealt{brandt19}, \citealt{bowler21}, \citealt{franson22}).
In this way these two methods complement each other to probe stellar and 
substellar masses at moderate to wide separations.\footnote{
TYC 3797-01268-1 has $Gaia$ RUWE = 2.81, consistent with its
binary status as discussed by \cite{jorissen20}, but it does not have an 
HGCA measurement.}

Figure~\ref{fig15} shows the $Gaia$ RUWE values and the 
SNR values of the HGCA acceleration measurements as a function of \he\
$\lambda$10830~\AA\ reduced widths.
Altogether 275 stars have $Gaia$ RUWE measurements.  
Among these, 207 stars have weak helium absorption (\rwhe~$<$~$-$4.85), 
24 of which have elevated RUWE values above 1.4. 
This implies a binary occurrence rate of at least 11.9$^{+2.4}_{-2.2}$\% 
using a binomial likelihood function.  
68 stars have strong helium absorption (\rwhe~$>$~$-$4.85); 17 of these have 
elevated RUWE values, indicating a 25.5$^{+5.5}_{-5.0}$\% binary 
occurrence rate.
Similarly, 99 bright stars from our program are in the HGCA catalog.  
For the 68 helium-weak sub-sample of stars, 11 (16.8$^{4.8}_{4.2}$\%) have 
SNR values above 3 and therefore are likely to have long-term astrometric 
accelerators (and hence companions).  
For the helium-strong sample, 11 stars (36.1$^{+8.9}_{-8.1}$\%) have SNR 
values above 3.  
These binomial posterior probability distributions for the companion 
frequencies are shown in Figure~\ref{fig16}.

In both cases $-$ stars with RUWE values and those with HGCA measurements
$-$ the binary occurrence rate is higher for stars with strong helium 
absorption compared to stars with weak helium.  
For the RUWE sub-sample, the significance of the difference
is at the $\approx$2.3-$\sigma$ level.  
For the HGCA sub-sample, the difference is at the $\approx$2.0-$\sigma$ level.
This suggests that there is likely (although not definitively) a relationship 
between the strength of the 10830~\AA \ helium line and stellar multiplicity.
Depending on the companion mass, separation, and eccentricity, it is possible 
that tidal interactions could impact the stellar chromosphere near periastron.
Alternatively, if enhanced helium is related to the engulfment of planets, the 
presence of an outer stellar or substellar companion could indicate that
the inner object underwent high-eccentricity tidal migration in the same way 
that hot Jupiters can form around main sequence stars (\citealt{fabrycky07}; 
\citealt{dawson18}).  
In this scenario, Kozai-Lidov oscillations with an outer inclined companion 
can pump the eccentricity of the inner object until the tidal influence
of the host star dampens, circularizes, and decays the planet's orbit.  
These hypotheses can be tested with a suite of follow-up observations including
rotation period measurements of the strong helium stars, radial velocity 
monitoring, and high-contrast imaging.  
$Gaia$'s final data release will provide additional 
clues about the nature of companions in these systems.

\vspace*{0.2in}
\subsection{The motions of Li-poor, He-strong Giants}\label{lipoorhestrong}

90\% of Li-poor program stars exhibit weak \species{He}{i} $\lambda$10830 
features.
Here we briefly comment on the 10\% of Li-poor stars that have strong 
$\lambda$10830 lines.
There are 16 stars in this category.
Six of them have astrometric indicators suggestive of the presence of
companion objects.
All of these stars have $Gaia$ RUWE~$>$~1.4 or HGCA~$>$~3, 
and four of them have HGCA~$>$~3.
Two additional stars, TYC~3304-00101-1 and TYC~3667-01280-1, are known to 
host giant planets (\citealt{niedzielski07}, \citealt{niedzielski16}).
And star TYC~3318-00020-1 is a rapid rotator (\vsini~=~17~\kmsec).
Thus 9 out of 16 Li-poor stars with strong \species{He}{i} lines are
known or suspected to have active chromospheres that may be due to present
or past binary interactions.

The other seven stars in this group have no known observational anomalies.
It is worth noting that none of these stars has a substantial literature
history; their only high-resolution spectroscopic studies appear to be
from \cite{adamow14} or \cite{afsar18b}.
They are worth further study, but we conclude here that the few red giants 
that are Li-poor but have strong $\lambda$10830 lines often can be
understood as resulting from binary interactions.

\vspace*{0.2in}
\section{CONCLUSIONS}\label{conclusions}

This paper has explored the relationship between Li abundances and 
absorption strengths of \species{He}{i} $\lambda$10830 in red giant stars.
Our survey considers only observational aspects of the Li-He linkage,
from which a few general conclusions may be drawn:
\begin{itemize}

\item 90\% of Li-poor stars, defined for this paper as those with 
\eps{Li}~$<$~1.25, have weak $\lambda$10830 transitions, \rwhe~$<$~$-$4.85.
In contrast, $\simeq$55\% of Li-rich stars have strong $\lambda$10830
absorptions.

\item The vast majority of He-strong stars reside on the He-burning RC and RHB.

\item Both Li-rich and He-strong stars are heavily concentrated in the Galactic
thin disk based on kinematics.

\item Many of the Li-rich and He-strong stars are also rapid rotators, or have
suspected binary companions. 

\item About half of the Li-poor, He-strong stars also have evidence for 
binarity or rapid rotation.

\item From these observational indicators we suggest, in agreement with previous
studies, that red giants are most likely to exhibit both high Li abundances
and strong \species{He}{i} $\lambda$10830 absorption lines if they are
He core-burning RC/RHB stars, and have current and/or past binary companions.

\end{itemize}
Several followup studies are underway to clarify the evolutionary
histories of red giants with strong $\lambda$10830 absorption lines.
The most immediate one is a large-sample survey of Kepler-field 
red giants.
The asteroseismic parameters known for these stars will identify the role of 
horizontal-branch ``age'' or time since the helium flash on the Li-He surface
combination.
In Table~\ref{tab-stars} we include 13 Kepler giants observed in 2021, but
our sample has now grown to 55 stars. 
Mallick \etal\ (in preparation) will report the \he\ $\lambda$10830 data for 
this larger sample of Kepler giants with known Li abundances, and discuss the 
evolutionary implications of these stars.

We are undertaking a field star survey based only on red giant 
rotation to estimate the importance of red giant envelope angular momentum in
creating the large $\lambda$10830 lines seen in many of our program stars.
It will also be worthwhile in the future to investigate $\lambda$10830 
strengths and Li abundances in binary-suspect giants, those with high values 
of RUWE and/or HGCA (Figure~\ref{fig15}).
Finally it would be good to conduct a southern-hemisphere extension to the 
present survey, which has been limited to $\delta$~$>$~$-$10$^{\circ}$.
That study could be undertaken with a southern-hemisphere high-resolution 
1$\mu$m instrument, such as WINERED \citep{ikeda22}, VLT/CRIRES 
\citep{kaeufl04,dorn14}, or PHOENIX \citep{hinkle03}.

\begin{acknowledgments}

We thank Claudia Aguilere-G{\' o}mez, Bengt Gustafsson, Noriyuki Matsunaga, 
George Preston, and our referee for helpful comments on this work.
These results are based on observations obtained with the Habitable-zone 
Planet Finder Spectrograph on the Hobby-Eberly Telescope. 
We thank the Telescope Operators at the HET for the 
skillful execution of our observations with HPF. 
The Hobby-Eberly Telescope is a joint project of the University of Texas at 
Austin, the Pennsylvania State University, Ludwig-Maximilians-Universität 
M{\" u}nchen, and Georg-August Universit{\" a}t Gottingen. 
The HET is named in honor of its principal benefactors, William P. Hobby and 
Robert E. Eberly. 
The HET collaboration acknowledges the support and resources from the Texas 
Advanced Computing Center.
We are happy to acknowledge support from NSF grant AST-1616040 (CS), 
and Technological Research Council of Turkey (TÜBİTAK), project No. 112T929
(MA).

\end{acknowledgments}

\facility{HET (HPF)}

\software{linemake (https://github.com/vmplacco/linemake), 
MOOG (Sneden 1973), 
IRAF (Tody 1986, Tody 1993), 
SPECTRE (Fitzpatrick \& Sneden 1987), 
Goldilocks (https://github.com/grzeimann/Goldilocks\_Documentation)}


\begin{center}
\begin{deluxetable}{lcrrrrrcrc}
\tablewidth{0pt}
\tablecaption{Program Star Data\label{tab-stars}}
\tablecolumns{10}
\tablehead{
\colhead{Star}                           &
\colhead{HD or BD}                       &
\colhead{V\tablenotemark{a}}             &
\colhead{B-V\tablenotemark{a}}           &
\colhead{V-J\tablenotemark{a}}           &
\colhead{p\tablenotemark{b}}             &
\colhead{\eps{Li}}                       &
\colhead{Source}                         &
\colhead{$EW_{\rm He}$}                  &
\colhead{broadening\tablenotemark{c}}    \\
\colhead{}                               &
\colhead{}                               &
\colhead{}                               &
\colhead{}                               &
\colhead{}                               &
\colhead{mas}                            &
\colhead{}                               &
\colhead{}                               &
\colhead{m\AA}                           &
\colhead{\kmsec}
}
\startdata
TYC 0014-00882-1 & BD+04 112 &  9.89 &  1.21 & 2.17 & 0.819 & $-$0.2 &  2 & 115 & 0.25 \\
TYC 0036-01321-1 &  HD 12203 &  6.76 &  1.02 & 1.73 & 5.999 &    2.0 &  4 &  80 & 0.27 \\
TYC 0037-00427-1 &  HD 12513 &  7.55 &  0.94 & 1.58 & 4.150 & $-$1.1 &  1 &  50 & 0.28 \\
TYC 0074-01190-1 &  HD 26573 &  6.57 &  0.91 & 1.60 & 8.728 &    1.4 &  1 & 551 & 0.32 \\
TYC 0096-00109-1 & BD+06 750 &  9.85 &  1.03 & 1.83 & 2.032 &    0.3 &  2 & 303 & 0.29 \\
TYC 0096-00163-1 &  \nodata  & 10.43 &  1.08 & 1.94 & 1.699 &    0.8 &  2 & 100 & 0.27 \\
TYC 0096-00301-1 &  \nodata  &  9.89 &  1.14 & 2.05 & 1.647 & $-$0.2 &  2 &  85 & 0.25 \\
TYC 0096-00378-1 &  \nodata  & 10.36 &  0.92 & 2.03 & 1.410 &    0.3 &  2 &  72 & 0.28 \\
TYC 0096-00659-1 &  HD 30897 &  8.26 &  1.28 & 2.02 & 3.666 &    1.0 &  2 & 125 & 0.24 \\
TYC 0096-00708-1 &  \nodata  & 10.24 &  1.02 & 1.89 & 1.470 &    0.3 &  2 &  70 & 0.28 \\
\enddata

\tablenotetext{a}{Magnitudes and colors are taken from SIMBAD \citep{wenger00}}
\tablenotetext{b}{Parallaxes are from $Gaia$ EDR3 \citep{GAIA21}}
\tablenotetext{c}{If the value is $<$1~\kmsec\ then the synthetic spectrum
                  broadening function is a Gaussian with the listed FWHM.
                  If it is $>$1~\kmsec\ then the estimated rotational \vsini\
                  is listed.  See the text for further comments.}
\tablerefs{{1} Bozkurt \etal\ to be surmitted,
           (2) \cite{adamow14},
           (3) \cite{mott17},
           (4) \cite{kumar11},
           (5) \cite{brown89b},
           (6) \cite{wallerstein82},
           (7) \cite{adamow14},
           (8) \cite{balachandran00},
           (9) \cite{hanni84},
           (10) \cite{deepak19},
           (11) \cite{adamow18},
           (12) \cite{luck82},
           (13) \cite{singh19b},
           (14) \cite{zhou18},
           (15) \cite{reddy16},
           (16) \cite{delaverny03},
           (17) \cite{yan18},
           (18) \cite{singh19a},
           (19) \cite{singh21},
           (20) \cite{bizyaev10},
           (21) \cite{carlberg12},
           (22) \cite{costa15},
           (23) \cite{yan21},
           (24) Nagarajan \etal\ to be submitted}
\vspace*{0.05in}
(This table is available in its entirety in machine-readable form.)
\end{deluxetable}
\end{center}

\begin{center}
\begin{deluxetable}{lrrl}
\tablewidth{0pt}
\tablecaption{Rotation Rates: This Study and Literature\label{tab-rotate}}
\tablecolumns{4}
\tablehead{
\colhead{Star}                           &
\colhead{{\vsini}\tablenotemark{a}}      &
\colhead{{\vsini}\tablenotemark{b}}      &
\colhead{Source}                         \\
\colhead{}                               &
\colhead{\kmsec}                         &
\colhead{\kmsec}                         &
\colhead{}           
}
\startdata
 TYC 0347-00762-1  &  18.0  &   15.3  &  \cite{bizyaev10}       \\
 TYC 0429-02097-1  &  12.0  &   11.3  &  \cite{yan18}           \\
 TYC 0455-02910-1  &  50.0  &     58  &  \cite{lyubimkov12}     \\
 TYC 0575-00918-1  &  35.0  &     42  &  \cite{kriskovics14}    \\
 TYC 1005-00073-1  &  80.0  &   84.1  &  \cite{massarotti08}    \\
 TYC 1395-02327-1  &  90.0  &     90  &  \cite{uesugi70}        \\
 TYC 2120-00320-1  &  32.0  &   28.7  &  \cite{takeda17}        \\
 TYC 2527-02031-1  &   7.8  &    6.1  &  \cite{rebull15}        \\
 TYC 2724-02354-1  & 140.0  &  139.7  &  \cite{massarotti08}    \\
 TYC 3134-00265-1  &  12.0  &  13.09  &  \cite{ceillier17}      \\
 TYC 3282-02270-1  &   9.0  &    7.2  &  \cite{goncalves20}     \\
 TYC 3318-00020-1  &  17.0  &   16.3  &  \cite{adamow14}        \\
 TYC 3340-01195-1  &   9.5  &    8.4  &  \cite{rebull15}        \\
 TYC 3590-03350-1  &   8.5  &    6.4  &  \cite{demedeiros02}    \\
 TYC 3676-02387-1  &  11.0  &   11.8  &  \cite{adamow14}        \\
 TYC 3797-01268-1  &  19.0  &   17.6  &  \cite{balachandran00}  \\
 TYC 4222-01254-1  &  30.0  &   22.1  &  \cite{guillout09}      \\
 TYC 4977-01458-1  &  17.0  &   15.5  &  \cite{massarotti08}    \\
\enddata

\tablenotetext{a}{this study}
\tablenotetext{b}{literature}

\end{deluxetable}
\end{center}

\end{document}